\documentclass[letter,11pt,DIV=12,abstract=true,numbers=noenddot,titlepage=false,twocolumn=false,draft=false]{scrartcl}
\pdfoutput=1 

\makeatletter
\DeclareOldFontCommand{\rm}{\normalfont\rmfamily}{\mathrm}
\DeclareOldFontCommand{\sf}{\normalfont\sffamily}{\mathsf}
\DeclareOldFontCommand{\tt}{\normalfont\ttfamily}{\mathtt}
\DeclareOldFontCommand{\bf}{\normalfont\bfseries}{\mathbf}
\DeclareOldFontCommand{\it}{\normalfont\itshape}{\mathit}
\DeclareOldFontCommand{\sl}{\normalfont\slshape}{\@nomath\sl}

\usepackage[usenames,dvipsnames]{color}
  \definecolor{hgreen}{rgb}{0,.3,0}
  \definecolor{hred}{rgb}{.3,0,0}
  \definecolor{hblue}{rgb}{0,0,.3}
  \definecolor{LightGray}{gray}{0.95}
  \definecolor{gray}{gray}{0.6}
\usepackage[
	    colorlinks=true,
	    linkcolor=hblue,
	    citecolor=hgreen,
	    filecolor=hblue,
	    urlcolor=hred
	    ]{hyperref}
\usepackage{mathrsfs}
\usepackage[intlimits]{amsmath}
\usepackage{amssymb}
\usepackage{slashed}
\usepackage[titletoc,title]{appendix}
\usepackage[affil-it]{authblk}
\usepackage{libertine}
\usepackage[numbers,sort&compress]{natbib}
\usepackage{graphicx}
\usepackage{physics}
\usepackage{bbm}
\usepackage{subcaption}

\setcapindent{1em}
\setkomafont{captionlabel}{\bfseries}
\setkomafont{caption}{\itshape}

\definecolor{Blu}{rgb}{0.,0.,1.}
\definecolor{Red}{rgb}{1.,0.,0.}
\definecolor{Green}{rgb}{0.,1.,0.}
\definecolor{Purple}{rgb}{0.5,0.,0.5}

\newcommand{\MS}{$\overline{\text{MS}}$}

\newcommand{\xt}{x_{t}}

\newcommand{\xtz}{x}

\newcommand{\mt}{m_t}

\newcommand{\Sc}{\mathscr{S}}

\allowdisplaybreaks  

\begin{document}
\renewcommand\Authands{, }

\title{\boldmath 
        Two-loop Electroweak Corrections to the Top-Quark Contribution to $\epsilon_K$
}

\date{\today}
\author{Joachim~Brod%
        \thanks{\texttt{joachim.brod@uc.edu}}}
\author{Sandra~Kvedarait\.e%
        \thanks{\texttt{kvedarsa@ucmail.uc.edu}}}
\author{Zachary~Polonsky%
        \thanks{\texttt{polonsza@mail.uc.edu}}}
	\affil{{\large Department of Physics, University of Cincinnati, Cincinnati, OH 45221, USA}}

\maketitle

\begin{abstract}
The parameter $\epsilon_K$ measures $CP$ violation in the neutral kaon
system. It is a sensitive probe of new physics and plays a prominent
role in the global fit of the Cabibbo-Kobabyashi-Maskawa matrix. The
perturbative theory uncertainty is currently dominated by the
top-quark contribution. Here, we present the calculation of the full
two-loop electroweak corrections to the top-quark contribution to
$\epsilon_K$, including the resummation of QED-QCD logarithms. We
discuss different renormalization prescriptions for the electroweak
input parameters. In the traditional normalization of the weak
Hamiltonian with two powers of the Fermi constant $G_F$, the top-quark
contribution is shifted by $-1\%$.
\end{abstract}
\setcounter{page}{1}

\section{Introduction}\label{sec:introduction}

The parameter $\epsilon_K$ describes CP violation in the neutral kaon
system and is one of the most sensitive probes of new physics. It can
be defined as~\cite{Proceedings:2001rdi}
\begin{equation}\label{eq:ek:def}
  \epsilon_K \equiv e^{i\phi_\epsilon} \sin\phi_\epsilon \frac{1}{2}
  \arg \bigg( \frac{-M_{12}}{\Gamma_{12}} \bigg)\,.
\end{equation}
Here, $\phi_\epsilon = \arctan(2\Delta M_K/\Delta\Gamma_K)$ where $\Delta M_K$ 
and $\Delta\Gamma_K$ are the mass and lifetime differences of the weak eigenstates 
$K_L$ and $K_S$. $M_{12}$ and $\Gamma_{12}$ are
the Hermitian and anti-Hermitian parts of the Hamiltonian that
determines the time evolution of the neutral kaon system. The
so-called short-distance contributions to $\epsilon_K$ enter
the matrix element $M_{12} = - \langle K^0 | \mathcal{L}^{\Delta S
  = 2}_{f=3}| \bar K^0 \rangle / (2\Delta M_K)$, up to higher powers
in the operator-product expansion.

Experimentally, $\epsilon_K$ is well-known, with absolute value
$|\epsilon_K| = (2.228 \pm 0.011)$ and an uncertainty at the permil
level~\cite{ParticleDataGroup:2020ssz}. The standard model (SM)
contributions to neutral kaon mixing are conveniently described by the
effective $|\Delta S| = 2$ Lagrangian with three active quark flavors,
\begin{equation}\label{eq:LS2:qcd}
\begin{split}
	\mathcal{L}^{|\Delta S|=2}_{f=3} = - \frac{G_F^2 M_W^2}{4 \pi^2}
    \big[ & \lambda_u^2 C_{S2}^{\prime \prime uu}(\mu) + \lambda_t^2 C_{S2}^{\prime \prime tt}
    (\mu) + \lambda_u \lambda_t C_{S2}^{\prime \prime ut}(\mu) \big]  Q_{S2}^{\prime \prime}
      + \textrm{h.c.} + \dots \,,
\end{split}
\end{equation}
valid at scales around $\mu = 2\,$GeV. Here,
\begin{equation}\label{eq:def:QS2}
Q_{S2}^{\prime \prime} =
\big(\overline{s}_L^\alpha \gamma_{\mu} d_L^\alpha\big) \otimes
\big(\overline{s}_L^\beta \gamma^{\mu}d_L^\beta\big)\,
\end{equation}
is the local $|\Delta S| = 2$ operator, where $\alpha$ and $\beta$ are
color indices, and the ellipsis denotes contributions of higher
dimension local operators and non-local contributions of $|\Delta S| =
1$ operators. The reason for the appearance of the double primes will
become clear later. The elements of the Cabibbo-Kobayashi-Maskawa
(CKM) matrix are combined into the parameters $\lambda_i \equiv
V_{is}^* V_{id}$. The long-distance SM contributions are comprised by
the hadronic matrix element of the local $|\Delta S| = 2$ operator,
and are given in terms of the kaon bag parameter $B_K =
0.7625(97)$~\cite{Aoki:2019cca}. Long-distance contributions that are
not included in $B_K$ are parameterized by the correction factor
$\kappa_\epsilon = 0.94(2)$~\cite{Buras:2010pza}.

The short-distance contributions are contained in the Wilson
coefficients $C_{S2}^{\prime \prime ij}$. In the SM, the leading-order
(LO) contributions to the Wilson coefficients are given by one-loop
box diagrams~\cite{Inami:1980fz}. Higher-order QCD corrections have
been calculated in renormalization-group (RG) improved perturbation
theory, using a slightly different representation of the $|\Delta S| =
2$ effective Lagrangian~\cite{Buchalla:1995vs}, in terms of
$\lambda_c$ and $\lambda_t$. However, it was shown
recently~\cite{Brod:2019rzc} that the
parameterization~\eqref{eq:LS2:qcd} in terms of $\lambda_u$ and
$\lambda_t$ leads to a perturbative QCD uncertainty smaller by an
order of magnitude. We will adopt this parameterization in this paper.

QCD corrections to the Wilson coefficients are known at
next-to-leading order (NLO) for $C_{S2}^{\prime \prime
  tt}$~\cite{Buras:1990fn} with a remaining uncertainty at the percent
level, and at next-to-next-to-leading order (NNLO) for $C_{S2}^{\prime
  \prime ut}$ with a remaining uncertainty below one
percent~\cite{Herrlich:1993yv, Brod:2010mj, Brod:2019rzc}. The
coefficient $C_{S2}^{\prime \prime uu}$ is also known at NNLO in
QCD~\cite{Brod:2011ty, Brod:2019rzc}, but has no effect on
$\epsilon_K$ (it does contribute to the neutral kaon mass difference).

In anticipation of the NNLO QCD calculation of $C_{S2}^{\prime \prime
  tt}$~\cite{BGSY} it is worthwhile to consider also the electroweak
corrections to the effective Lagrangian~\eqref{eq:LS2:qcd}. Without an
explicit calculation, the electroweak renormalization scheme of the
input parameters is left unspecified and amounts to an uncertainty of
the order of a few percent -- which can no longer be neglected at the
current level of precision.

The electroweak corrections to $C_{S2}^{\prime \prime tt}$ could, in
principle, be adapted from the calculation of the corresponding
corrections for $B$-meson mixing presented in
Ref.~\cite{Gambino:1998rt}. However, in our opinion, a reconsideration
of the old calculation is worthwhile for a number of reasons. First,
the application to $\epsilon_K$ involves a lower energy scale than the
one relevant in $B$-meson mixing; accordingly, the QED-QCD resummation
of the appearing logarithms might be necessary. Second, we will
discuss the scheme dependence of the corrections in detail, since the
usual proofs of scheme independence~\cite{Buchalla:1995vs} fail in our
case; this topic also has not been addressed before. Last, our
calculation presents the first independent check of the results in
Ref.~\cite{Gambino:1998rt}.

In this work, we calculate the coefficient $C_{S2}^{\prime \prime
  tt}$, proportional to $\lambda_t^2$, to NLO in the electroweak
interactions. This fixes the renormalization scheme of the electroweak
input parameters contained in the prefactor $G_F^2 M_W^2$. In fact,
when only considering QCD effects, there are several equivalent ways
of rewriting the prefactor, using the tree-level relation
\begin{equation}\label{eq:GF:tree}
G_F
= \frac{\pi \alpha}{\sqrt{2} M_W^2 s_w^2} \,,
\end{equation}
where $\alpha = e^2/(4\pi)$ the electromagnetic
coupling\footnote{Throughout this paper, the electromagnetic coupling
  is understood to be fixed in the five-flavor scheme at scale $\mu =
  M_Z$, unless otherwise noted.}  and $s_w^2 = \sin^2\theta_w$ with
the weak mixing angle $\theta_w$. Essentially, this choice specifies
which experimental data are used as parametric input for our
prediction. The numerical difference between the different schemes is
expected to be large at LO, as exemplified by the 5\% difference
between the on-shell and \MS{} values of $s_w^2$.

For our analysis, it is useful to write the effective
Lagrangian in the following form:
\begin{equation}\label{eq:LS2:g2}
\mathcal{L}^{|\Delta S|=2}_{f=3} = \lambda_t^2 c_{S2}^{tt} (\mu)
Q_{S2}^{(\prime, \prime\prime)} + \textrm{h.c.} + \dots \,,
\end{equation}
where the ellipsis now also includes contributions not proportional to
$\lambda_t^2$. Using the tree-level relation~\eqref{eq:GF:tree}, we
express $c_{S2}^{tt}$ in three different ways:
\begin{equation}\label{eq:cS2}
c_{S2}^{tt} (\mu) = - \frac{2\pi^2}{M_W^2 s_w^4}
C_{S2}^{tt} (\mu)\,,
\qquad
c_{S2}^{tt} (\mu) = - \frac{G_F}{\sqrt{2} s_w^2}
C_{S2}^{\prime tt}\,,
\qquad
c_{S2}^{tt} (\mu) = - \frac{G_F^2 M_W^2}{4 \pi^2}
C_{S2}^{\prime \prime tt}\,.
\end{equation}
As explained in Sec.~\ref{sec:muon}, this effectively absorbs
different parts of the radiative corrections into the measured value
of the muon decay rate. In the first and second relation in
Eq.~\eqref{eq:cS2} we have factored out the powers $(\alpha/(4\pi))^2$
and $\alpha/(4\pi)$, respectively, and absorbed them into rescaled
operators, defined as
\begin{equation}\label{eq:def:QS2:A2}
Q_{S2} =
\bigg(\frac{\alpha}{4\pi}\bigg)^2
\big(\overline{s}_L^\alpha \gamma_{\mu} d_L^\alpha\big) \otimes
\big(\overline{s}_L^\beta \gamma^{\mu}d_L^\beta\big)\,,
\end{equation}
and
\begin{equation}\label{eq:def:QS2:GF}
Q_{S2}' =
\frac{\alpha}{4\pi}
\big(\overline{s}_L^\alpha \gamma_{\mu} d_L^\alpha\big) \otimes
\big(\overline{s}_L^\beta \gamma^{\mu}d_L^\beta\big)\,,
\end{equation}
while $Q_{S2}''$ has been defined in Eq.~\eqref{eq:def:QS2}. With
these conventions, the RG evolution is described by the same anomalous
dimension in all three cases, see Sec.~\ref{sec:qcd}, and the LO
values
\begin{equation}
C_{S2}^{tt} (\mu) = C_{S2}^{\prime tt} (\mu) = C_{S2}^{\prime \prime tt} (\mu) = S(x_t)
\end{equation}
coincide. They are given by the modified~\cite{Brod:2019rzc} Inami-Lim
box function~\cite{Inami:1980fz}
\begin{equation}
\Sc(x_t) = \frac{4 x_t - 11 x_t^2 + x_t^3}{4(x_t-1)^2}
         + \frac{3 x_t^3}{2(x_t-1)^3} \log x_t \,,
\end{equation}
where $x_t \equiv m_t^2/M_W^2$, and we neglect a tiny correction of
${\cal O}(m_c^2/M_W^2) \sim 10^{-4}$~\cite{Brod:2019rzc}. We will
refer to the first normalization in Eq.~\eqref{eq:cS2} as ``A2'', the
second as ``GF'', and the third as ``GF2''. While, at LO in the weak
interactions, the three parameterizations are equivalent, the
numerical prediction depends on the chosen normalization and the
scheme of the input parameters. These arbitrary dependences will be
mitigated to a large degree by the NLO electroweak corrections. The
aim of this paper is to show this explicitly, and to provide an
updated numerical prediction for $\epsilon_K$ including the
electroweak corrections.

This paper is organized as follows. In Sec.~\ref{sec:match:ew} we
define the effective Lagrangian and provide details on our two-loop
calculation. Sec.~\ref{sec:ew} contains the discussion of the various
electroweak renormalization schemes and the error estimate on the
electroweak corrections. In Sec.~\ref{sec:qcd} we include the effects
of the strong interaction and discuss the scheme dependence of our
result. We conclude in Sec.~\ref{sec:conclusions}. A number of
appendices contains more details of our work. In App.~\ref{sec:ren} we
collect the explicit counterterms that were needed in the
renormalization of the SM amplitude. The necessary counterterms in the
effective theory are collected in App.~\ref{sec:z}. We give a formal
proof of the scheme independence of the top-quark contribution to the
$|\Delta S| = 2$ amplitude including QED effects in
App.~\ref{sec:scheme}. Finally, the full analytic two-loop result is
presented in App.~\ref{sec:full}.

\section{Calculation of the two-loop electroweak matching corrections}\label{sec:match:ew}

Our strategy is to perform the matching calculation using the ``A2''
normalization in the \MS{} scheme. We can then easily change to
different normalization conventions and renormalization schemes for
the input parameters; this will be discussed in detail in this
section. The variation of the NLO results between the different
prescriptions is used as one way to estimate the remaining theory
uncertainty.

\subsection{Matching calculation}

We write the physical five-flavor effective Lagrangian, obtained by
integrating out the top quark together with the weak gauge bosons $W$
and $Z$ as well as the Higgs boson, in the form\footnote{The RG
  evolution to the three-flavor effective Lagrangian will be
  considered in Sec.~\ref{sec:qcd}.}
\begin{equation}\label{eq:LS2:A2}
\mathcal{L}^{|\Delta S|=2}_{f=5} = - \frac{2\pi^2}{M_W^2 s_w^4} \lambda_t^2
C_{S2}^{tt} (\mu)
Q_{S2} + \textrm{h.c.} + \dots \,.
\end{equation}
The single physical operator contributing to our results has been
defined in Eq.~\eqref{eq:def:QS2:A2}. Our definition of the evanescent
operators is given in App.~\ref{sec:z}, thus fixing the
renormalization scheme of our results.

The initial condition for the Wilson coefficient $C_{S2}^{tt}$ at the
electroweak scale is calculated from the difference of the $|\Delta S|
= 2$ amplitudes in the full SM and the five-flavor EFT. We expand the
Wilson coefficients in the five-flavor EFT as
\begin{equation}\label{eq:expand:C}
  C_{S2}^{tt}
= C_{S2}^{tt,(0)}
  + \frac{\alpha^{(5)}}{4\pi} C_{S2}^{tt,(se)} + \ldots \,,
\end{equation}
where the superscript ``(5)'' emphasizes that the electroweak coupling
is defined in the five-flavor scheme. We expand the amplitude in the
full SM as
\begin{equation}\label{eq:A:SM}
{\mathcal A}^{tt}
 = - \frac{2\pi^2}{M_W^2 s_w^4} \lambda_t^2
   \sum_i \bigg( {\mathcal A}_{i}^{tt,(0)}
   + \frac{\alpha^{(6)}}{4\pi} {\mathcal A}_{i}^{tt,(se)} + \ldots
   \bigg) \langle Q_{i} \rangle^{(0)} \,,
\end{equation}
in terms of the six-flavor coupling. This implies in the matching we
have to include the finite threshold correction~\cite{Bobeth:2013tba}
\begin{equation}
  \alpha^{(6)} = \alpha^{(5)} \bigg[ 1 - \frac{\alpha^{(5)}}{4\pi}
                \bigg( \frac{2}{3} + 14 \log\frac{\mu}{M_W}
                       - \frac{32}{9} \log\frac{\mu}{m_t} \bigg) \bigg] \,,
\end{equation}
in order to express both results in terms of $\alpha^{(5)}$. The sum
in Eq.~\eqref{eq:A:SM} runs over all operators in the basis.

The LO matching conditions are simply ${\mathcal A}_{i}^{tt,(0)} =
C_{i}^{tt,(0)}$, and we find the following non-zero contributions (the
terms of order $\epsilon$ in the physical Wilson coefficient are
relevant for the two-loop calculation)
\begin{align}
\begin{split}
C_{S2}^{tt,(0)} & = \frac{4 x_t - 11 x_t^2 + x_t^3}{4(x_t-1)^2}
               + \frac{3 x_t^3}{2(x_t-1)^3} \log x_t \\
             & + \epsilon \bigg[ 
                 \frac{4 x_t - 11 x_t^2 + x_t^3}{4(x_t-1)^2}
               + \frac{3 x_t^3}{2(x_t-1)^3} \log x_t 
                 \bigg] \log\Big(\frac{\mu^2}{M_W^2}\Big) \\
             & + \epsilon \bigg[ 
                 \frac{3 x_t^3 - 41 x_t^2 - 12 x_t}{8(x_t - 1)^2}
               - \frac{x_t^4 - 14 x_t^3 - 8 x_t - 4 x_t}{4(x_t - 1)^3} \log x_t
               - \frac{3 x_t^3}{4(x_t - 1)^3}\log^2 x_t \bigg] \\
             & - 4 \epsilon C_{E_{S2}}^{tt,(0)} \,,
\end{split}\\
C_{E_{S2}}^{tt,(0)} & = - \frac{x_t + x_t^2}{4(x_t-1)^2}
               + \frac{x_t^2}{2(x_t-1)^3} \log x_t \,.
\end{align}
We obtained this result by equating the amplitudes in the full SM and
the five-flavor EFT defined above. We set all external momenta and
fermion masses apart from the top quark to zero and employ dimensional
regularization in $d=4-2\epsilon$ dimensions for both ultraviolet (UV)
and infrared (IR) divergences.

\begin{figure}[t]
        \centering
        \includegraphics[width=\textwidth]{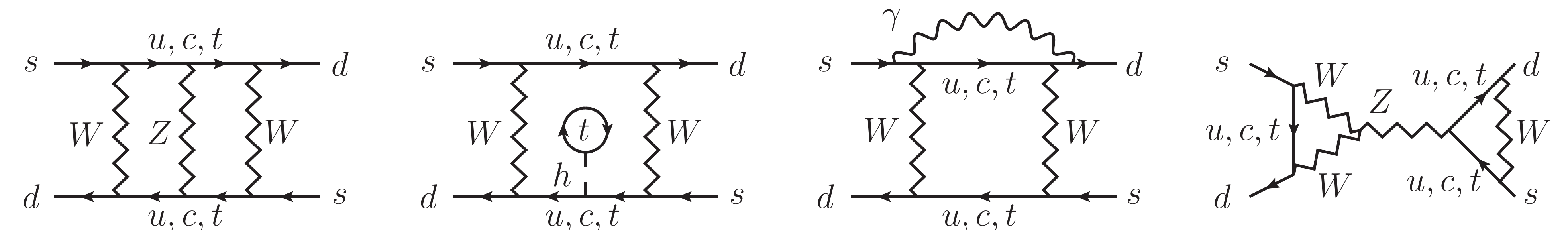}
        \caption{Sample two-loop Feynman diagrams contributing to the
          initial condition of $C_{S2}^{\prime \prime tt}$.
	\label{fig:2loop}}
\end{figure}

\begin{figure}[t]
        \centering
        \includegraphics[width=0.5\textwidth]{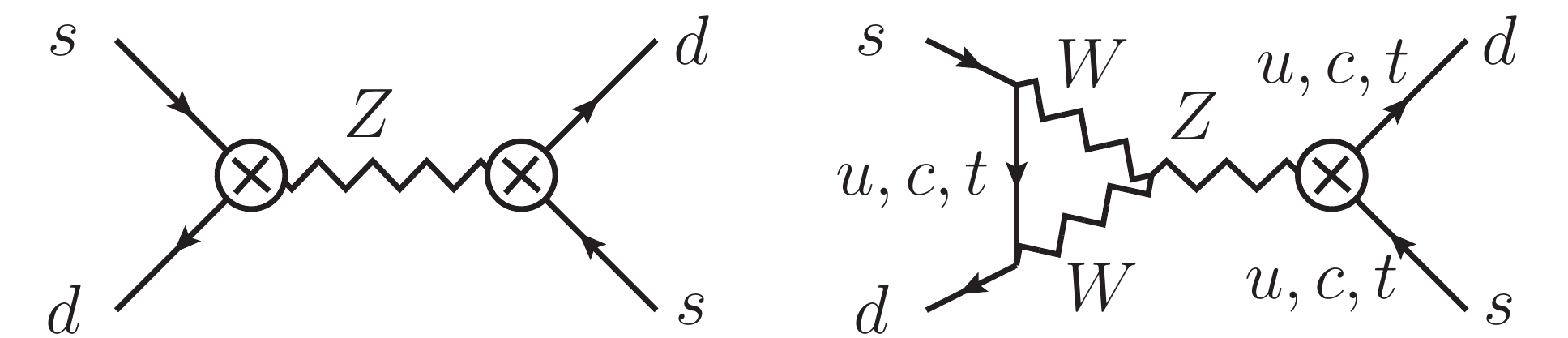}
        \caption{Sample Feynman diagrams contributing to the
          off-diagonal field renormalization: ${\mathcal
            A}_{Q_{S2}''}^{tt,(se),\text{tree-level Z}}$ (left panel),
          and ${\mathcal A}_{Q_{S2}''}^{tt,(se),\text{one-loop Z}}$
          (right panel). The circled cross denotes a counterterm
          insertion.
	\label{fig:countt}}
\end{figure}

In the remainder of this section, we discuss the matching at NLO in
the electroweak interactions. We split the NLO SM amplitude into
several contributions as follows:
\begin{equation}
\begin{split}
{\mathcal A}_{Q_{S2}''}^{tt,(se)}
=
{\mathcal A}_{Q_{S2}''}^{tt,(se),\text{two-loop}}
+
{\mathcal A}_{Q_{S2}''}^{tt,(se),\text{countt}}
+
{\mathcal A}_{Q_{S2}''}^{tt,(se),\text{one-loop Z}}
+
{\mathcal A}_{Q_{S2}''}^{tt,(se),\text{tree-level Z}} \,.
\end{split}
\end{equation}
Here, ${\mathcal A}_{Q_{S2}''}^{tt,(se),\text{two-loop}}$ denotes the
contribution of the ${\mathcal O}(30\,000)$ genuine two-loop diagrams
(see Fig.~\ref{fig:2loop} for sample Feynman diagrams). The
counterterm contributions arising from the renormalization of the
parameters in the LO result are
\begin{equation}
\begin{split}
{\mathcal A}_{Q_{S2}''}^{tt,(se),\text{countt}}
 =  
    \bigg[ \bigg(   4 \frac{\delta e}{e}
                  - 2 \frac{\delta s_w^2}{s_w^2}
                  - \frac{\delta M_W^2}{M_W^2} \bigg) {\mathcal A}_{Q_{S2}''}^{tt,(0)}
    + \delta m_t \frac{\partial {\mathcal A}_{Q_{S2}''}^{tt,(0)}}{\partial m_t}
    + \delta M_W^2 \frac{\partial {\mathcal A}_{Q_{S2}''}^{tt,(0)}}{\partial M_W^2}
    \bigg] \,.
\end{split}
\end{equation}
The explicit forms of the renormalization constants are given in
App.~\ref{sec:ren}. We have included all tadpole graphs in both the
full two-loop amplitude and the counterterms, such that the
renormalization constants are gauge independent. The contributions of
external field renormalization are comprised by ${\mathcal
  A}_{Q_{S2}''}^{tt,(se),\text{tree-level Z}}$ and ${\mathcal
  A}_{Q_{S2}''}^{tt,(se),\text{one-loop Z}}$ (see
Fig.~\ref{fig:countt} for sample diagrams, and App.~\ref{sec:ren} for
the explicit form of the field renormalization constants). We include
finite matching contributions in the renormalization of the external
fields, such that the fields are normalized in the same way in the
full and effective theories (cf. Ref.~\cite{Buras:2002vd}).

As an additional check, we calculated all one-loop diagrams with
single counterterm insertions, including those corresponding to
unphysical Goldstone boson field and mass renormalizations as well as
Goldstone-$W$ mixing. After a careful application of the
Slavnov-Taylor Identities~\cite{Gambino:1998rt}, we analytically
verified that the two methods of calculating ${\mathcal
  A}_{Q_{S2}''}^{tt,(se),\text{countt}}$ are identical.

Parameterizing the matrix elements in the EFT as
\begin{equation}
  \langle Q_i \rangle = \sum_j r_{ij} \langle Q_j \rangle^{(0)} \,,
\end{equation}
where $\langle Q_j \rangle^{(0)}$ denotes the tree-level matrix
element, and writing
\begin{equation}
  r_{ij} = \delta_{ij} + \frac{\alpha^{(5)}}{4\pi} r_{ij}^{(se)} + \ldots \,,
\end{equation}
the NLO matching condition is
\begin{equation}
  C_{i}^{tt,(se)}
= {\mathcal A}_{i}^{tt,(se)}
  - \sum_j C_{j}^{tt,(0)}  \Big( r_{ji}^{(se)} + Z_{ji}^{(se)} \Big) \,,
\end{equation}
where $Z_{ji}^{(se)}$ denotes the renormalization constants for the
Wilson coefficients; their explicit form is given in
App.~\ref{sec:z}. The final result for $C_{i}^{tt,(se)}$ is given in
App.~\ref{sec:full}.

All diagrams have been calculated using self-written
\texttt{FORM}~\cite{Vermaseren:2000nd} routines, implementing the
two-loop recursion presented in Refs.~\cite{Davydychev:1992mt,
  Bobeth:1999mk}. The amplitudes were generated using
\texttt{qgraf}~\cite{Nogueira:1991ex}. As a check on our calculation,
we verified that both IR and UV divergences cancel in the matching,
yielding a finite result. We checked analytically that the full
amplitude is independent of the matching scale to the considered order
in the electroweak interaction.

\subsection{Normalization to muon decay}\label{sec:muon}

When considering the prediction of low-energy observables at higher
orders in the electroweak interactions, it is often advantageous to
employ the muon lifetime as one of the experiments fixing the values
of the input parameters. By a suitable normalization of the results,
this allows for the absorption of parts of the radiative corrections
into the muon lifetime measurement~\cite{Sirlin:1981ie}.

In the context of our calculation, we regard the Fermi constant $G_F$
as the Wilson coefficient of muon decay in the Fermi effective
theory~\cite{vanRitbergen:1999fi}. We write the effective Lagrangian
as
\begin{equation}
  {\mathcal L}_F
= - \frac{4}{\sqrt{2}} \frac{\pi\alpha}{\sqrt{2}M_W^2 s_w^2} G_\mu Q_\mu + \text{H.c.} \,,
\end{equation}
where
\begin{equation}
  Q_\mu = (\bar\nu_{\mu,L} \gamma^\mu \mu_L)
         (\bar\nu_{e,L} \gamma_\mu e_L) \,,
\end{equation}
and we expand
\begin{equation}
  G_\mu = G_\mu^{(0)} + \frac{\alpha}{4\pi} G_\mu^{(se)} + \ldots \,.
\end{equation}
The LO matching then just yields $G_\mu^{(0)} = 1$, while a one-loop
matching calculation gives
\begin{equation}\begin{split}\label{eq:Gmu:1loop}
 G^{(se)}_\mu =&
 \frac{144 x^2 z^2 - 18 x y z^2 - 21 y^2 z^2 + 15 y z^2 + 14 y z + 16 y
       - 12 z^2 - 24}{24 y (z-1)}\\
 & - \frac{22 y^2 z^3 - 53 y^2 z^2 + 28 y^2 z - 7 y z^2 + 38 y z - 28 y
          - 12 z + 12}{4 y (z-1)^2 (y z-1)} \log(z)\\
 & + \frac{ - 108 x^2 z^2 + 27 x y z^2 - 32 y z + 32 y}{18 y (z - 1)} \log(x)
 + \frac{3 y^2 z^3 - 6 y z^2}{4 (z-1) (y z-1)} \log(y)\\
 & + \frac{216 x^2 z^2 - 54 x y z^2 - 27 y^2 z^2 + 27 y z^2 - 134 y z + 
         188 y - 54 z^2 - 108}{36 y (z - 1)} \log\Big(\frac{\mu^2}{M_Z^2}\Big)\,,
\end{split}\end{equation}
with $x \equiv m_t^2/M_Z^2$, $y \equiv M_h^2/M_Z^2$, and $z =
M_Z^2/M_W^2$. (Note that, again, a finite threshold correction has
been included in order to express the result in terms of the
five-flavor $\alpha$.) This result corresponds to the following
definition of the relevant evanescent operator:
\begin{equation}
  E_\mu = (\bar\nu_{\mu,L} \gamma^{\mu_1} \gamma^{\mu_2} \gamma^{\mu_3} \mu_L)
         (\bar\nu_{e,L} \gamma_{\mu_1} \gamma_{\mu_2} \gamma_{\mu_3} e_L)
         - (16 - 4\epsilon) Q_\mu \,.
\end{equation}
This form of the evanescent operator ensures the validity at
${\mathcal O}(\alpha)$ of the Fierz relations that have been used to
calculate the QED corrections to the muon decay matrix element in the
Fermi theory~\cite{vanRitbergen:1999fi}.

We can now discuss the different normalizations introduced in
Sec.~\ref{sec:introduction}, including the NLO corrections. For ``GF''
normalization we have
\begin{equation}
  c_{S2}^{tt} = - \frac{1}{\sqrt{2} s_w^2} G_F C_{S2}^{\prime tt}
            = - \frac{2\pi^2}{M_W^2 s_w^4} C_{S2}^{tt} \,.
\end{equation}
Inserting the all-order relation $G_F = \pi\alpha/(\sqrt{2}M_W^2s_w^2)
\times G_\mu$ and expanding $G_\mu$, we find
\begin{equation}
  \frac{\alpha}{4\pi} C_{S2}^{\prime tt}
=   C_{S2}^{tt,(0)}
  + \frac{\alpha}{4\pi} 
    \bigg[   C_{S2}^{tt,(se)}
           - C_{S2}^{tt,(0)} G_\mu^{(se)} \bigg]
  + \ldots \,.
\end{equation}
Similarly, for ``GF2'' normalization we have
\begin{equation}
  c_{S2}^{tt} = - \frac{M_W^2}{4\pi^2} G_F^2 C_{S2}^{\prime \prime tt}
            = - \frac{2\pi^2}{M_W^2 s_w^4} C_{S2}^{tt} \,,
\end{equation}
or 
\begin{equation}
  \bigg(\frac{\alpha}{4\pi}\bigg)^2 C_{S2}^{\prime \prime tt}
=   C_{S2}^{tt,(0)}
  + \frac{\alpha}{4\pi} 
    \bigg[   C_{S2}^{tt,(se)}
           - 2 C_{S2}^{tt,(0)} G_\mu^{(se)} \bigg]
  + \ldots \,.
\end{equation}
(Recall that we have absorbed factors of $\alpha/(4\pi)$ into the
definitions of the operators.) These relations allow us to obtain
$C_{S2}^{\prime tt}$ and $C_{S2}^{\prime \prime tt}$ from our explicit
calculations of $C_{S2}^{tt}$ and $G_\mu$.

\section{Discussion of electroweak renormalization schemes}\label{sec:ew}

The purpose of this section is to identify the residual theory
uncertainty with regard to the higher-order electroweak
corrections. We will estimate the uncertainty by studying different
renormalization schemes for the input parameters, and the residual
matching scale dependence in the \MS{} scheme. The immediate problem
facing us is the dominant residual scale dependence with regard to
QCD. In order to isolate the electroweak effects, we will therefore
completely ignore QCD in this section. We define a formally scale- and
scheme-independent quantity by multiplying the Wilson coefficient with
the matrix element of the $Q_{S2}$ operator. This serves to cancel the
part of the scale dependence related to the anomalous dimension of the
Wilson coefficient, and to cancel its residual scheme dependence (see
the detailed discussion in Sec.~\ref{sec:qcd}). We do not include any
other finite terms in the matrix element. In principle, this matrix
element could be calculated, including QCD effects, using lattice
methods. The absolute value of the matrix element plays no role in our
numerics, as in this section we are only interested in the scale and
scheme variation, not the absolute values for the Wilson
coefficients. These will be considered, including QCD, in the next
section. To be specific, we multiply the Wilson coefficient by the
partonic matrix element
\begin{equation}\label{eq:ME}
  \langle Q_{S2} \rangle
  \equiv \bigg[ 1 + \frac{\alpha}{4\pi} r^{(se)} + \ldots \bigg]
  \langle Q_{S2} \rangle^{(0)} \,,
\end{equation}
where $\langle Q_{S2} \rangle^{(0)}$ denotes the tree-level matrix
element, and the LO QED correction is given by
\begin{equation}
  r^{(se)} = \frac{1}{9} a_{11} - \frac{4}{3} \log\frac{\mu_t}{\mu_\text{had}}\,,
\end{equation}
with $\mu_\text{had} = 2\,$GeV.

\begin{table}
  \centering
  \begin{tabular}{|l|l|}
    \hline
    $M_Z = 91.1876(21)\,$GeV & $M_t^\text{pole} = 172.5(7)\,$GeV\\\hline
    $M_W = 80.379(12)\,$GeV & $\sin^2\theta_w^{\overline{\text{MS}}} = 0.23141(4)$\\\hline
    $\alpha_s(M_Z) = 0.1179(10)$ & $\alpha(M_Z)^{-1} = 127.952(9)$\\\hline
    $G_F = 1.1663787(6) \times 10^{-5}\,$GeV$^{-2}$ & $M_h = 125.25(17)\,$GeV \\\hline
  \end{tabular}
  \caption{Primary input values used in our numerics; all numbers are
    taken from the PDG~\cite{ParticleDataGroup:2020ssz}.
    \label{tab:input}}
\end{table}

\subsection{\MS{} scheme}\label{sec:msbar}

Our primary input parameters in the \MS{} scheme are
\begin{equation}
  \alpha_s(M_Z)\,, \quad G_F\,, \quad \alpha(M_Z)\,, \quad M_Z\,, \quad M_t\,, \quad M_h\,.
\end{equation}
Their values are collected in Tab.~\ref{tab:input}. In the \MS{}
scheme it is most convenient to express all physical quantities in
terms of the running parameters
\begin{equation}\label{eq:coup:msbar}
  g_1\,, \quad g_2\,, \quad g_s\,, \quad y_t\,, \quad \lambda\,, \quad v\,.
\end{equation}
To determine the initial conditions of these couplings at $\mu = M_Z$,
we first use RunDec~\cite{Chetyrkin:2000yt} to convert the top-quark
pole mass to QCD-\MS{} using three-loop accuracy. We find $mt(mt) =
162.9(7)\,$GeV. We then convert the (electroweak) on-shell masses
$M_Z$, $m_t(m_t)$, $M_h$ to \MS{} masses, using the one-loop
expressions from~\cite{Jegerlehner:2001fb, Jegerlehner:2002em}, and
choosing the matching scale to be the respective on-shell masses. We
take into account the correction~\eqref{eq:Gmu:1loop} when converting
$G_F$ to $v$. We then fit the initial conditions of the
couplings~\eqref{eq:coup:msbar} such that they reproduce the central
values of the converted input parameters at their respective scales,
employing two-loop running of the couplings. Note that the strong
coupling $g_s$ is neglected throughout this section (apart from the
conversion of the top-quark mass to QCD-\MS{}.)

The two-loop beta functions for the SM are taken from
Ref.~\cite{Arason:1991ic}. However, their result for the running of
the Higgs vacuum expectation value is not consistent with our
treatment of tadpole graphs (in fact, the result for $\beta_v$ given
in Ref.~\cite{Arason:1991ic} is only valid in Landau gauge). A
gauge-independent result consistent with our conventions can be
extracted from Refs.~\cite{Jegerlehner:2001fb,
  Jegerlehner:2002em}. For convenience, we collect the corresponding
beta function here, writing $\beta_v = \beta_{v}^{(1)} +
\beta_{v}^{(2)} + \ldots\,$. We find
\begin{align}\label{eq:beta:v}
 \beta_{v}^{(1)} & = \frac{1}{16\pi^2}
               \bigg( - \frac{3}{2} \lambda
                      + \frac{3}{4} g_1^2
                      + \frac{9}{4} g_2^2
                      - \frac{3g_1^4}{4\lambda}
                      - \frac{3g_1^2g_2^2}{2\lambda}
                      - \frac{9g_2^4}{4\lambda}
                      - 3 y_t^2
                      + 12 \frac{y_t^4}{\lambda} \bigg) \,,
\end{align}
where we counted the contributions of colored fermions in
Ref.~\cite{Jegerlehner:2002em} with multiplicity three. As a check, we
obtained the identical result by an explicit calculation of the
one-loop renormalization of the Higgs mass, and using the beta
function for the Higgs quartic from Ref.~\cite{Arason:1991ic} (see
also Ref.~\cite{Brod:2020lhd}). As a further consistency check, we
note that $\beta_{v}^{(1)}$ is identical to the coefficient of the
$\log(\mu^2)$ term in Eq.~\eqref{eq:Gmu:1loop}, after expressing the
mass ratios in terms of couplings. Also the two-loop beta function can
be extracted from Refs.~\cite{Jegerlehner:2001fb, Jegerlehner:2002em},
as a combination of the beta functions for the $W$ mass and the $g_2$
gauge coupling. This yields
\begin{equation}
\begin{split}
  \beta_{v}^{(2)}
& = \frac{1}{(16\pi^2)^2}
    \bigg(   \frac{1}{32} g_2^4
           - \frac{63}{16} g_1^2 g_2^2
           - \frac{701}{96} g_1^4
           + \frac{379}{24} \frac{g_1^6}{\lambda}
           + \frac{559}{24} \frac{g_1^4 g_2^2}{\lambda}\\
& \hspace{5.5em}
           + \frac{289}{24} \frac{g_1^2 g_2^4}{\lambda}
           - \frac{305}{8} \frac{g_2^6}{\lambda}
           - \frac{3}{2} g_1^2 \lambda
           - \frac{9}{2} g_2^2 \lambda
           + \frac{63}{8} \lambda^2 \\
& \hspace{5.5em}
           - \frac{85}{24} g_1^2 y_t^2
           - \frac{45}{8} g_2^2 y_t^2
           - \frac{21}{4} y_t^4
           + \frac{19}{2} \frac{g_1^4 y_t^2}{\lambda} \\
& \hspace{5.5em}
           - 21 \frac{g_1^2 g_2^2 y_t^2}{\lambda}
           + \frac{9}{2} \frac{g_2^4 y_t^2}{\lambda}
           + \frac{16}{3} \frac{g_1^2 y_t^4}{\lambda}
           - 60 \frac{y_t^6}{\lambda}
           + 9 y_t^2 \lambda
           - 12 g_2^2 g_s^2
    \bigg) \,.
\end{split}
\end{equation}

\begin{figure}[t]
        \centering
        \includegraphics[width=\textwidth]{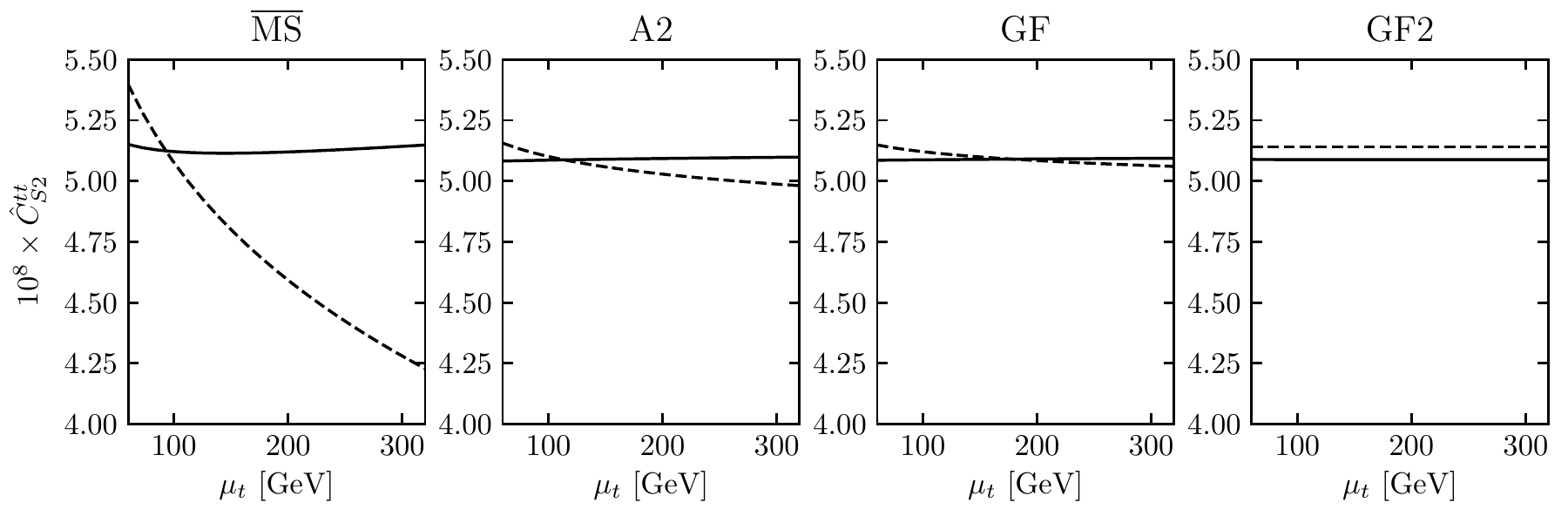}
        \caption{Residual electroweak matching scale dependence in the
          \MS{} and hybrid schemes. The dashed line shows the LO
          result, while the solid line show the NLO result. See text
          for details.
	\label{fig:all:schemes}}
\end{figure}

Part of the matching scale dependence of the Wilson coefficient cancels
the scale dependence of the running parameters in the prefactor of the
Lagrangian. Hence, we define a formally scale and scheme independent
Wilson coefficient in the \MS{} scheme (where we always use ``A2''
normalization)
\begin{equation}
  \hat C_{S2}^{tt} =
  \frac{\alpha^2(\mu)}{8 m_W^2(\mu) \big(s_w^{\overline{\text{MS}}}(\mu)\big)^4}
  C_{S2}^{tt}(\mu) \langle Q_{S2}\rangle (\mu)
\end{equation}
where $\alpha$, $m_W$, and $s_w^{\overline{\text{MS}}}$ are defined in
the \MS{} scheme, and the matrix element is defined in
Eq.~\eqref{eq:ME}. Varying the matching scale $\mu_t$ between
$60\,$GeV and $320\,$GeV, we see that the matching scale dependence is
reduced from $\pm 12\%$ at LO to $\pm 0.4\%$ at NLO (see
Fig.~\ref{fig:all:schemes}, left panel).

\subsection{On-shell scheme}

Here, we define all parameters in the on-shell scheme (OS) regarding
the electroweak interaction. (We always treat the top-quark mass as an
\MS{} mass regarding the strong interaction.) The weak mixing angle is
defined, in the on-shell scheme, by $\sin^2\theta_w^\text{OS} = 1 -
M_W^2/M_Z^2$. The $W$-boson mass itself is a function of the primary
input parameters; to obtain its numerical value $M_W = 80.354\,$GeV,
we use the results of Ref.~\cite{Awramik:2003rn}. Again, we define a
formally scale- and scheme-independent Wilson coefficient. Its
explicit form depends on the chosen normalization and is given by
\begin{equation}
  \hat C_{S2}^{tt} =
  \frac{\alpha^2(\mu)}{8 M_W^2 \big(s_w^{\text{OS}}\big)^4}
  C_{S2}^{tt}(\mu) \langle Q_{S2}\rangle (\mu)
\end{equation}
for ``A2'' normalization, by
\begin{equation}
  \hat C_{S2}^{tt} =
  \frac{\alpha(\mu) G_F}{4\sqrt{2} \big(s_w^{\text{OS}}\big)^2}
  C_{S2}^{\prime tt}(\mu) \langle Q_{S2}\rangle (\mu)
\end{equation}
for ``GF'' normalization, and by
\begin{equation}
  \hat C_{S2}^{tt} =
  \frac{G_F^2 M_W^2}{4\pi^2}
  C_{S2}^{\prime \prime tt}(\mu) \langle Q_{S2}\rangle (\mu)
\end{equation}
for ``GF2'' normalization. Here, $s_w^{\text{OS}}$ and $M_W$ are
defined in the on-shell scheme, while $\alpha$ is (always) defined as
a \MS{} coupling (note that the dependence on $s_w^{\text{OS}}$ drops
out in GF2 normalization).

\begin{figure}[t]
        \centering
        \includegraphics[width=\textwidth]{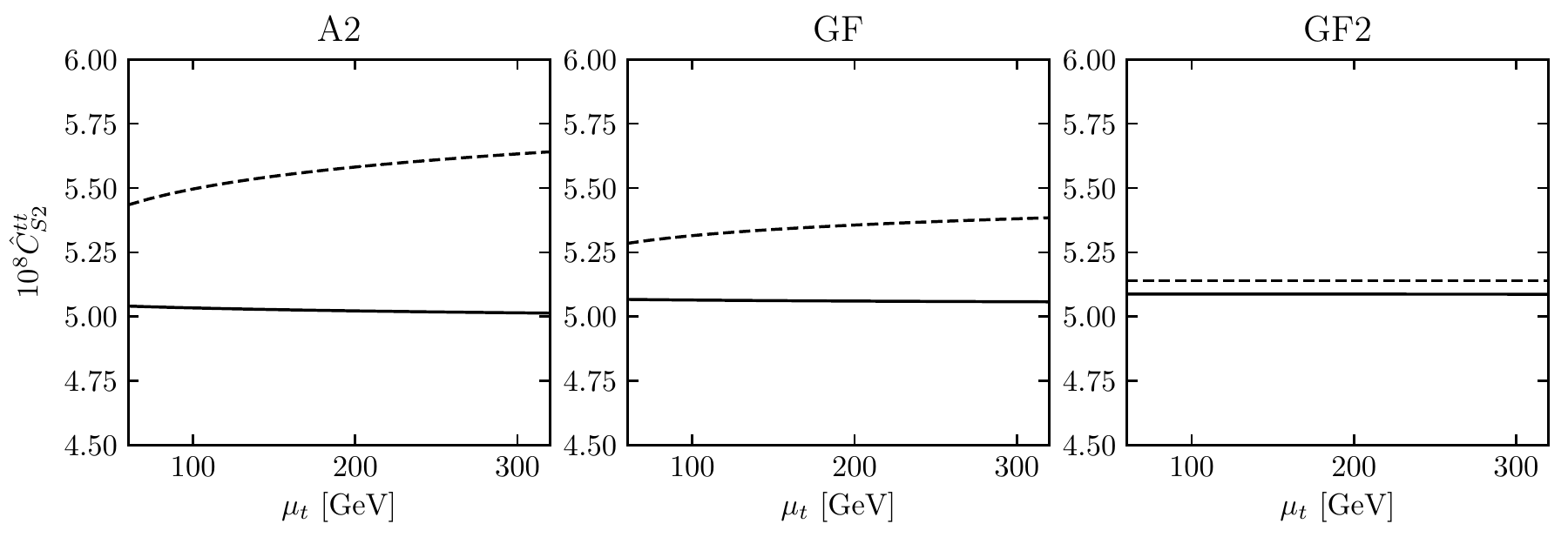}
        \caption{Residual electroweak matching scale dependence in the
          full OS scheme. The dashed line shows the LO result, while
          the solid line show the NLO result. See text for details.
	\label{fig:os}}
\end{figure}

As is clearly visible in Fig.~\ref{fig:os}, the NLO corrections in the
on-shell scheme are large, indicating slow convergence of the
perturbation series. As discussed in Ref.~\cite{Bobeth:2013tba}, this
can be attributed to the large top-mass dependence of the on-shell
counterterm for the weak mixing angle. Hence, the on-shell scheme is
not suitable for low-energy observables like $\epsilon_K$, and we will
not use it in our error estimate.

\subsection{Hybrid scheme}

In this scheme, we define all masses in the on-shell scheme regarding
the electroweak interaction, while the weak mixing angle that appears
in the prefactor is defined in the \MS{} scheme:
$\sin^2\theta_w^{\overline{\text{MS}}} = g_1^2 / (g_1^2 +
g_2^2)$. Accordingly, we choose a different set of primary input
parameters for the numerics in this scheme, namely,
\begin{equation}
  \alpha_s(M_Z)\,, \quad \sin^2\theta_w\,, \quad \alpha(M_Z)\,, \quad M_W\,, \quad M_t\,, \quad M_h\,.
\end{equation}
Again, we fit the initial conditions of the \MS{}
parameters~\eqref{eq:coup:msbar} as described in Sec.~\ref{sec:msbar},
neglecting the strong interaction. We define formally scale- and
scheme-independent Wilson coefficients in this scheme as follows. For
``A2'' normalization,
\begin{equation}
  \hat C_{S2}^{tt} =
  \frac{\alpha^2(\mu)}{8 M_W^2 \big(s_w^{\overline{\text{MS}}}(\mu)\big)^4}
  C_{S2}^{tt}(\mu) \langle Q_{S2}\rangle (\mu) \,,
\end{equation}
while for ``GF'' normalization,
\begin{equation}
  \hat C_{S2}^{tt} =
  \frac{\alpha(\mu) G_F}{4\sqrt{2} \big(s_w^{\overline{\text{MS}}}(\mu)\big)^2}
  C_{S2}^{\prime tt}(\mu) \langle Q_{S2}\rangle (\mu) \,.
\end{equation}
Here, $s_w^{\overline{\text{MS}}}(\mu)$ is defined in the \MS{}
scheme, while $M_W$ is still defined in the on-shell scheme. Note that
the hybrid scheme coincides with the on-shell scheme for the ``GF2''
normalization, up to the tiny numerical difference in the $M_W$ mass.

The Wilson coefficients for the three different normalizations are
shown in Fig.~\ref{fig:all:schemes} (right three panels). While the LO
results in ``A2'' and ``GF'' normalization still show sizeable scale
dependence, all NLO curves are essentially flat. As an additional
estimate of the unknown higher-order electroweak effects, we take half
of the absolute value of the difference between the largest and the
smallest NLO value. Again, this translates into a $\pm 0.4\%$
uncertainty.

\section{Inclusion of QCD}\label{sec:qcd}

As is well-known, the dominant corrections to the LO SM prediction of
the $|\Delta S| = 2$ weak Lagrangian arise from QCD and have to be
taken into account in the final numerics. In this section, we will
summarize the status of the QCD corrections and combine them with our
electroweak corrections.

The large separation between the electroweak scale, $\mu_t \sim m_t,
M_W$ and the scale $\mu_\text{had} \sim 2\,$GeV where the hadronic
matrix elements are evaluated mandate the use of RG-improved
perturbation theory, summing powers of $\alpha_s
\log(\mu_t/\mu_\text{had})$ to all orders. Here, we include also the
(small) QED corrections. The formalism is
well-known~\cite{Buchalla:1995vs, Buras:1993dy}. We now expand the
Wilson coefficient, including the QCD terms, as
\begin{equation}\label{eq:expand:C}
  C_{S2}^{tt}(\mu)
= C_{S2}^{tt,(0)}(\mu)
  + \frac{\alpha_s(\mu)}{4\pi} C_{S2}^{tt,(1)}(\mu)
  + \frac{\alpha}{\alpha_s(\mu)} C_{S2}^{tt,(e)}(\mu)
  + \frac{\alpha}{4\pi} C_{S2}^{tt,(se)}(\mu) + \ldots \,.
\end{equation}
The initial conditions $C_{S2}^{tt,(0)}(\mu_t)$ and $
C_{S2}^{tt,(se)}(\mu_t)$ have been given in the previous section. The
coefficient $C_{S2}^{tt,(e)}(\mu_t)$ does not receive a matching
contribution; it is purely generated by the RG flow and corresponds to
LL QED resummed logarithms of the form $\alpha \alpha_s^n
\log^{n+1}(\mu_t/\mu_\text{had})$. The last term contains the summed
NLL QED logarithms of the form $\alpha \alpha_s^n
\log^{n}(\mu_t/\mu_\text{had})$. The coefficient
$C_{S2}^{tt,(1)}(\mu_t)$ receives matching contributions from two-loop
box diagrams with gluon exchange. As a check of our setup we
calculated this term and find
\begin{equation}
\begin{split}
C_{S2}^{tt,(1)}(\mu)  
& =       \frac{8\pi^2}{9 x_t}
        + \frac{10 x_t^5 + 55 x_t^4 + 149 x_t^3 + 4 x_t^2}{6 (x_t-1)^4} \log^2(x_t) \\
& \quad + \frac{ - 15 x_t^5 - 2 x_t^4 - 543 x_t^3 + 388 x_t^2 - 148 x_t + 32}{6 (x_t-1)^4} \log(x_t) \\
& \quad + \frac{25 x_t^4 + 248 x_t^3 + 75 x_t^2 - 92 x_t + 32}{6 (x_t-1)^3} \\
& \quad + \text{Li}_2(1 - x_t) \frac{10 x_t^5 + 74 x_t^4 - 2 x_t^3 + 40 x_t^2 - 48 x_t + 16}{3x_t(x_t-1)^3} \\
& \quad + \frac{5 x_t^4 + 87 x_t^2 - 20 x_t}{2 (x_t-1)^3} \log(x_\mu)
        + \frac{3 x_t^4 - 39 x_t^3}{(x_t-1)^4}
          \log(x_t) \log(x_\mu) \,,
\end{split}
\end{equation}
where $x_\mu \equiv \mu^2/M_W^2$, and $\text{Li}_2 (x) = -\int_0^x du
\, \ln (1-u)/u$ is the usual dilogarithm. This expression is in full
agreement the well-known result presented in Ref.~\cite{Buras:1990fn}.

\begin{figure}[t]
        \centering
        \includegraphics[width=0.5\textwidth]{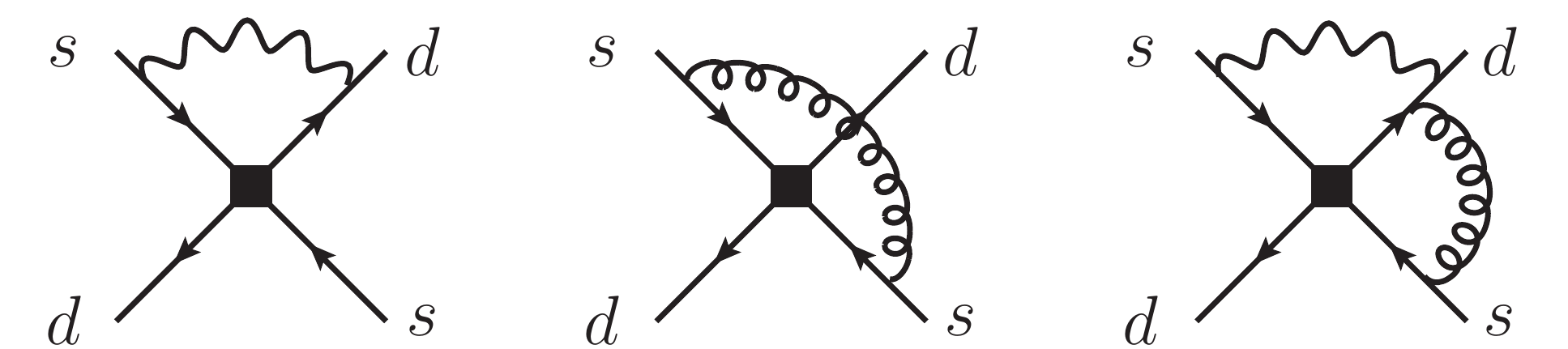}
        \caption{Sample Feynman diagrams contributing to the two-loop
          anomalous dimension.
	\label{fig:eff}}
\end{figure}

The RG evolution is most conveniently written in terms of an evolution
matrix, such that $C_{S2}^{tt}(\mu) = C_{S2}^{tt}(\mu_0)
U(\mu_0,\mu,\alpha)$. We expand
\begin{equation}\label{eq:U:expansion}
\begin{split}
U(\mu_0,\mu,\alpha) = 
   U^{(0)}(\mu_0,\mu)
 + \frac{\alpha_s(\mu)}{4\pi}U^{(1)}(\mu_0,\mu)
 + \frac{\alpha}{\alpha_s(\mu)}U^{(e)}(\mu_0,\mu)
 + \frac{\alpha}{4\pi}U^{(se)}(\mu_0,\mu)
\end{split}
\end{equation}
in terms of the couplings defined at the low scale $\mu$, and find the
following contributions to the Wilson coefficient at the low scale:
\begin{align}\label{eq:C:low:coefficients}
C_{S2}^{tt,(0)}(\mu) & = C_{S2}^{tt,(0)}(\mu_0) U^{(0)}(\mu_0,\mu)\,,\\
C_{S2}^{tt,(1)}(\mu) & = \eta C_{S2}^{tt,(1)}(\mu_0) U^{(0)}(\mu_0,\mu) + C_{S2}^{tt,(0)}(\mu_0) U^{(1)}(\mu_0,\mu)\,,\\ 
C_{S2}^{tt,(e)}(\mu) & = C_{S2}^{tt,(0)}(\mu_0) U^{(e)}(\mu_0,\mu) + \eta^{-1} C_{S2}^{tt,(e)}(\mu_0) U^{(0)}(\mu_0,\mu)\,,\\
\begin{split}
C_{S2}^{tt,(se)}(\mu) & = \eta C_{S2}^{tt,(1)}(\mu_0) U^{(e)}(\mu_0,\mu)
                 + \eta^{-1} C_{S2}^{tt,(e)}(\mu_0) U^{(1)}(\mu_0,\mu) \\
            & \quad + C_{S2}^{tt,(se)}(\mu_0) U^{(0)}(\mu_0,\mu)
                    + C_{S2}^{tt,(0)}(\mu_0) U^{(se)}(\mu_0,\mu)\,,
\end{split}
\end{align}
where we have introduced the ratio
$\eta=\alpha_s(\mu_0)/\alpha_s(\mu)$. The explicit expression for the
evolution matrix can be found in Ref.~\cite{Buras:1993dy} and involves
the anomalous dimension of the Wilson coefficients. It is given by
\begin{equation}\label{eq:adm}
  \gamma_{S2}^{(0)} =   \frac{\alpha_s}{\pi}
                    + \bigg(\frac{4N_f}{9} - 7\bigg) \frac{\alpha_s^2}{16\pi^2}
                    + \frac{\alpha}{3\pi}
                    -\frac{148}{9} \frac{\alpha\alpha_s}{16\pi^2}\,.
\end{equation}
The first two terms are well-known~\cite{Buchalla:1995vs}, while the
QED corrections are new. They have been calculated by extracting the
UV poles of the relevant one- and two-loop diagrams (see
Fig.~\ref{fig:eff} for examples) using the infrared rearrangement
described in Ref.~\cite{Chetyrkin:1997fm}.

The result~\eqref{eq:adm} is valid for the Wilson coeffients
($C_{S2}^{tt}$, $C_{S2}^{\prime tt}$, and $C_{S2}^{\prime \prime tt}$)
in all three normalizations conventions, due to our absorbing powers
of $\alpha$ into the definitions of the corresponding operators. Note
that the $\alpha_s^2$ term depends on the explicit form of the
evanescent operators, given in App.~\ref{sec:z}, while the other three
terms are scheme independent, as we have verified by explicit
calculation.

This last observation deserves further discussion. Our two-loop result
$C_{S2}^{tt,(se)}$ does depend on the definition of evanescent
operators in Eq.~\eqref{eq:evan:full}. The SM prediction for the
observable $\epsilon_K$ must, of course, be independent of such
arbitrary choices; in fact, the scheme dependence of the Wilson
coefficient will cancel exactly against the corresponding scheme
dependence of the hadronic matrix element (a proof is given in
App.~\ref{sec:scheme}.) In the literature on $\epsilon_K$, the scheme
independent product of Wilson coefficient and matrix element is
usually factorized into two {\em separately scheme- and
  scale-independent} quantities, namely, the QCD correction factors
$\eta_{tt}$ and $\eta_{ut}$, and the kaon bag factor $\hat B_K$. This
is achieved by writing the evolution matrix as $U(\mu_0,\mu,\alpha) =
K^{-1}(\mu_0,\alpha) U^{(0)}(\mu_0,\mu,\alpha) K(\mu,\alpha)$, and
combining the $K$ factors, together with the appropriate part of the
LO evolution matrix, with the Wilson coefficients and the matrix
elements to yield scheme-independent quantities (see
Refs.~\cite{Buras:1993dy, Buchalla:1995vs} for details).

In our case, this strategy fails when including QED corrections, as
the ${\mathcal O}(\alpha\alpha_s)$ anomalous dimension is scheme
independent by itself. This is consistent with the general expression
for the scheme dependence of anomalous dimensions given in
Ref.~\cite{Buras:1993dy} and App.~\ref{sec:scheme}, because here the
anomalous dimension is a one-dimensional matrix, i.e. just a number.
Therefore, the definition of the scheme-invariant correction factor
$\eta_{tt}$ cannot be extended to include QED effects (as a
fixed-order expansion in $\alpha$).

It follows that, in the absence of a determination of the hadronic
matrix element including QED corrections, our result will be scheme
dependent. However, the scheme dependence is tiny. Numerically, the
explicit dependence on the coefficient $a_{11}$, defined in
App.~\ref{sec:z}, is
\begin{equation}
  C_{S2}^{\prime \prime tt}(2\,\text{GeV}) = (3.90 - 0.0003 a_{11} ) \times 10^{-8} \,.
\end{equation}
This dependence will cancel once a full matrix element is
available. However, we expect the finite shift in the hadronic matrix
element to be equally tiny, of order $\alpha/(4\pi) \sim 10^{-4}$. The
bulk of the effect of electroweak corrections is captured by the
matching calculation at the weak scale, not by the QED effects in the
effective theory. We will therefore neglect the scheme dependence in
our numerical discussion.

\begin{figure}[t]
        \centering
        \includegraphics[width=\textwidth]{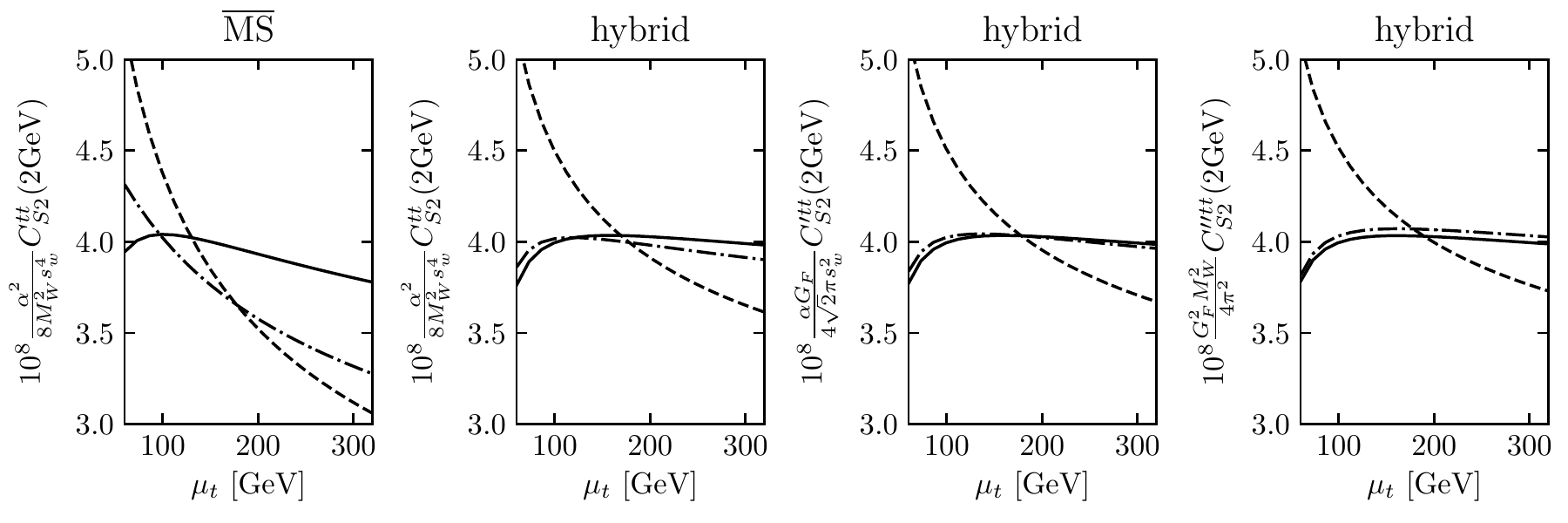}
        \caption{Residual matching scale dependence with full QCD
          resummation, in the \MS{} and hybrid schemes. The dashed
          line shows the LO result, the dash-dotted line shows the
          result including NLO QCD corrections, while the solid line
          show the full (QED and QED) NLO result. See text for
          details.
	\label{fig:qcd}}
\end{figure}

To quantify the impact of the electroweak corrections, we examine the
numerical values of the Wilson coefficients in the hybrid
renormalization scheme, as it shows the best convergence properties
and least residual scale dependence (cf. Fig.~\ref{fig:qcd}). The
values in the three different normalization conventions are given in
Tab.~\ref{tab:WC}, both for the NLL QCD result only, and including the
full NLO electroweak corrections (with NLL QED resummation). The shown
uncertainties are half of the difference between maximum and minimum
values when varying the matching scale $\mu_t$ between $60\,$GeV and
$320\,$GeV. We see that the shift from NLL QCD to full electroweak is
actually largest, $-1\%$, in the conventionally used ``GF2''
normalization, while in the ``A2'' and ``GF'' normalizations the shift
is $+0.5\%$ and $-0.5\%$, respectively.\footnote{Using the (even) more
  conservative range $40\,\text{GeV} \leq \mu_t \leq 320\,\text{GeV}$
  leads to a $-1\%$ shift in all three normalizations.} The resulting
NLO values are in perfect agreement.

\begin{table}
  \centering
  \begin{tabular}{|l|c|c|}
    \hline
     & NLL QCD & NLL QCD \& NLL QED \\\hline
     $ \alpha^2/(8M_W^2 s_w^4) C_{S2}^{tt}(2\,\text{GeV}) \times 10^{8}$ & $3.96(6)$ & $3.98(6)$ \\\hline
     $ \alpha G_F/(4\sqrt{2}\pi s_w^2) C_{S2}^{\prime tt}(2\,\text{GeV}) \times 10^{8}$ & $4.00(4)$ & $3.98(5)$ \\\hline
     $ G_F^2 M_W^2/(4\pi^2) C_{S2}^{\prime \prime tt}(2\,\text{GeV}) \times 10^{8}$ & $4.02(5)$ &$3.98(5)$ \\\hline
  \end{tabular}
  \caption{Wilson coefficients. Uncertainty given is scale variation only.
    \label{tab:WC}}
\end{table}


With the three-loop QCD corrections in the top sector and the
electroweak corrections in the charm-top sector still outstanding, we
refrain from an extensive numerical discussion of our results. The
shift in the Wilson coefficient is small, of order one percent, in the
hybrid scheme in all normalizations. Since both the residual scale
dependence and the spread between NLO values (considering electroweak
effects only) is both $\pm 0.4\%$ (see Sec.~\ref{sec:ew}), we suggest
the following temporary prescription that is most easily implemented:
Adopt the traditional ``GF2'' normalization, and multiply $\eta_{tt}$
by the electroweak correction factor $(1-\Delta_{tt})$, with
$\Delta_{tt} = 0.01 \pm 0.004$. With this prescription, the SM
prediction presented in Ref.~\cite{Brod:2019rzc} is shifted by
$-0.7\%$ to $|\epsilon_K| = 2.15(6)(7)(15) \times 10^{-3}$, with the
errors corresponding to short-distance, long-distance, and parametric
uncertainties (see Ref.~\cite{Brod:2019rzc} for details).

\section{Conclusions}\label{sec:conclusions}

We have presented the complete two-loop electroweak corrections to the
top-quark contribution to the parameter $\epsilon_K$. The analogous
electroweak corrections for $B^0 - \overline{B^0}$ mixing have been
presented previously in Ref.~\cite{Gambino:1998rt}; as a check, we
reproduced their numerical results using the input parameters given in
that reference. In our calculation we used three different
normalizations for the effective Lagrangian and several
renormalization schemes for the electroweak input parameters. While
these lead to different numerical predictions at LO, the final results
agree well if the electroweak corrections are taken into account. We
assess the theoretical uncertainty by studying the residual
electroweak matching scale dependence and the residual dependence on
the renormalization scheme for the input parameters, leading to a
error of $\pm 0.4\%$ asscociated with unknown higher-order
corrections.

We then considered the full RG evolution in the effective theory at
NLO in QCD and mixed QED-QCD. In particular, we discuss how the scheme
dependence, related to the freedom in choosing the evanescent
operators, cancels against the hadronic matrix element if evaluated
including QED corrections. While these corrections are currently
unknown, they are expected to be tiny, as is the leftover scheme
dependence of our result.

Numerically, the inclusion of the electroweak corrections amounts to a
$-1\%$ downward shift in the Wilson coefficient for the top-quark
contribution defined at the hadronic scale $\mu_\text{had} = 2\,$GeV,
when using the conventional normalization of the effective
Lagrangian. We defer a more extensive numerical study of the effect on
the SM prediction of $\epsilon_K$ to the future, when the three-loop
QCD corrections in the top sector~\cite{BGSY}, the two-loop
electroweak corrections in the charm-top sector~\cite{BKPY}, and
possibly an updated hadronic matrix element will be
available. Instead, we just shift the usual QCD correction factor
$\eta_{tt}$ by a factor $1-0.010(4)$. This shifts the SM prediction
presented in Ref.~\cite{Brod:2019rzc} by $-0.7\%$ to $|\epsilon_K| =
2.15(6)(7)(15) \times 10^{-3}$. The errors correspond to
short-distance, long-distance, and parametric uncertainties.

\section*{Acknowledgments}

We thank Jens Erler for useful communication regarding the electroweak
input parameters, and Martin Gorbahn and Emmanuel Stamou for many
discussions and comments on the manuscript. JB acknowledges support by
DOE grant DE-SC0011784. The Feynman diagrams have been generated using
\texttt{jaxodraw}~\cite{Binosi:2003yf}, based on
\texttt{axodraw}~\cite{Vermaseren:1994je}.

\appendix

\section{Mass and field renormalization constants}\label{sec:ren}

All mass renormalization constants include the tadpole contributions
and are thus gauge-parameter independent:
\begin{align}
  \begin{split}
  \delta M_W^2 & = \frac{e^2}{16\pi^2\epsilon} M_W^2
               \bigg[   \frac{3}{4}
                      - \frac{11}{12s_w^2}
                      + \frac{3}{4} \frac{s_w^2}{c_w^2}
                      - \frac{3}{2s_w^2} \frac{m_t^2}{M_W^2} \\
    & \hspace{6.4em}  - \frac{3}{4s_w^2} \frac{M_h^2}{M_W^2}
                      + \frac{6}{s_w^2} \frac{m_t^4}{M_W^2 M_h^2}
                      - \frac{3}{2s_w^2c_w^4} \frac{M_W^2}{M_h^2}
                      - \frac{3}{s_w^2} \frac{M_W^2}{M_h^2}
               \bigg] \,,
  \end{split}\\
  \begin{split}
  \delta M_Z^2 & = \frac{e^2}{16\pi^2\epsilon} M_Z^2
               \bigg[   \frac{29}{6c_w^2}
                      - \frac{11}{12} \frac{1}{s_w^2c_w^2}
                      + \frac{11}{3} \frac{s_w^2}{c_w^2}
                      - \frac{3}{2s_w^2c_w^2} \frac{m_t^2}{M_Z^2}\\
  &  \hspace{6.4em}   - \frac{3}{4s_w^2c_w^2} \frac{M_h^2}{M_Z^2}
                      + \frac{6}{s_w^2c_w^2} \frac{m_t^4}{M_Z^2 M_h^2}
                      - \frac{3}{2s_w^2c_w^2} \frac{M_Z^2}{M_h^2}
                      - \frac{3}{s_w^2} \frac{M_Z^2}{M_h^2}
                      + 3 \frac{M_Z^2}{M_h^2}
               \bigg] \,,
  \end{split}\\
  \delta M_h^2 & = \frac{e^2}{16\pi^2\epsilon} M_h^2
               \bigg[ - \frac{3}{4s_w^2c_w^2}
                      - \frac{3}{2s_w^2}
                      + \frac{3}{4s_w^2} \frac{M_h^2}{M_W^2}
                      + \frac{3}{2s_w^2} \frac{m_t^2}{M_W^2} \bigg] \,,\\
  \begin{split}
  \delta m_t & = \frac{e^2}{16\pi^2\epsilon} m_t
               \bigg[ - \frac{3}{4s_w^2c_w^2} \frac{M_Z^2}{M_h^2}
                      - \frac{3}{2s_w^2} \frac{M_W^2}{M_h^2}
                      - \frac{3}{8s_w^2} \frac{M_h^2}{M_W^2}\\
                      & \qquad \qquad \qquad
                      - \frac{1}{3c_w^2}
                      + \frac{3}{8s_w^2} \frac{m_t^2}{M_W^2}
                      + \frac{3}{s_w^2} \frac{m_t^4}{M_W^2 M_h^2} \bigg] \,.
  \end{split}
\end{align}
It follows that
\begin{equation}
\frac{\delta s_{w}^2}{s_w^2}
   = - \frac{c_w^2}{s_w^2} \frac{\delta c_w^2}{c_w^2}
   = \frac{e^2}{16\pi^2\epsilon} 
     \bigg[ \frac{11}{3} + \frac{19}{6s_w^2} \bigg]\,.
\end{equation}
Note that all tadpole contributions to $\delta s_{w}^2$ cancel. The
charge renormalization is given by
\begin{equation}
  \delta Z_{e} = \frac{e^2}{16\pi^2\epsilon}
               \bigg[ \frac{8N_u}{9} + \frac{2N_d}{9} + \frac{2N_\ell}{3} - \frac{7}{2} \bigg] \,.
\end{equation}
The divergent part of the one-loop off-diagonal field renormalization
constant (in 't~Hooft-Feynman gauge) is
\begin{align}\label{eq:Zsd:inf}
  \delta Z_{sd}^L & = - \frac{e^2}{16\pi^2\epsilon}
                     \frac{V_{ts} V_{td}^*}{s_w^2}
                     \frac{\xt}{4} \,.
\end{align}
The finite part of the one-loop off-diagonal field renormalization
constant (in 't~Hooft-Feynman gauge) is
\begin{equation}\label{eq:Zsd:fin}
  \delta Z_{sd}^L = - \frac{e^2}{16\pi^2} \frac{V_{ts} V_{td}^*}{s_w^2}
                   \bigg[   \frac{3(\xt^2 + \xt)}{8(\xt - 1)}
                          + \frac{\xt - 4\xt^2}{4(\xt - 1)^2} \log \xt
                          + \frac{\xt}{4} \log \frac{\mu^2}{\mt^2}
                   \bigg] \,,
\end{equation}
where $x_t \equiv m_t^2/M_W^2$. Both Eqs.~\eqref{eq:Zsd:inf}
and~\eqref{eq:Zsd:fin} are in agreement with the results in
Ref.~\cite{Gambino:1998rt}.
The divergent part of the one-loop diagonal field renormalization
constant for the down-type quarks (in 't~Hooft-Feynman gauge) is
\begin{align}
  \delta Z_{dd}^L & = - \frac{e^2}{16\pi^2\epsilon} \frac{1}{s_w^2}
                     \bigg[
                        V_{td} V_{td}^* \frac{\xt}{4}
                      + \frac{13}{18}
                      + \frac{1}{36c_w^2}
                     \bigg]\,.
\end{align}
The finite part of the one-loop diagonal field renormalization
constant for the down-type quarks (in 't~Hooft-Feynman gauge) is
\begin{equation}
\begin{split}
  \delta Z_{dd}^L & = - \frac{e^2}{16\pi^2} \frac{V_{td} V_{td}^*}{s_w^2}
                     \bigg[   \frac{3(\xt^2 + \xt)}{8(\xt - 1)}
                            + \frac{\xt - 4\xt^2}{4(\xt - 1)^2} \log \xt
                            + \frac{\xt}{4} \log \frac{\mu^2}{\mt^2}
                     \bigg]\\
                 & \quad + \frac{e^2}{16\pi^2 s_w^2}
                     \bigg[   \frac{11}{36}
                            + \frac{1}{72c_w^2}
                            + \frac{c_w^2}{18}
                            - \bigg( (1+c_w^2) \frac{1}{9} + \frac{1}{36c_w^2} \bigg)
                              \log \xtz
                            - \frac{1}{2} \log \xt
                     \bigg]\\
                 & \quad - \frac{e^2}{16\pi^2 s_w^2}
                     \bigg[   \frac{11}{18}
                            + \frac{1}{36c_w^2}
                            + \frac{c_w^2}{9}
                     \bigg]
                     \log\frac{\mu^2}{m_t^2} \,.
\end{split}
\end{equation}
The finite part differs from the expression in
Ref.~\cite{Gambino:1998rt} due to our using dimensional regularization
for the IR divergences.


\section{Renormalization constants in the effective theory}\label{sec:z}

The evanescent operators necessary for the calculation of the matching
conditions and anomalous dimensions up to two-loop in QCD and QED
are chosen as~\cite{Brod:2010mj}
\begin{align}\label{eq:evan:full}
F_{S2}^{(0)} & = \bigg(\frac{\alpha}{4\pi}\bigg)^2
\big(\overline{s}_L^\alpha \gamma_{\mu} d_L^\beta\big) \otimes
\big(\overline{s}_L^\beta \gamma^{\mu}d_L^\alpha\big) - Q_{S2}\,,
\\
E_{S2}^{(1)} & =  \bigg(\frac{\alpha}{4\pi}\bigg)^2
\big(\overline{s}_L^\alpha \gamma_{\mu_1} \gamma_{\mu_2} \gamma_{\mu_3} d_L^\alpha\big) \otimes
\big(\overline{s}_L^\beta \gamma^{\mu_1} \gamma^{\mu_2} \gamma^{\mu_3} d_L^\beta\big)
 - (16 - a_{11}\epsilon - 4\epsilon^2) Q_{S2} \,,
\\
F_{S2}^{(1)} & = \bigg(\frac{\alpha}{4\pi}\bigg)^2
\big(\overline{s}_L^\alpha \gamma_{\mu_1} \gamma_{\mu_2} \gamma_{\mu_3} d_L^\beta\big) \otimes
\big(\overline{s}_L^\beta \gamma^{\mu_1} \gamma^{\mu_2} \gamma^{\mu_3} d_L^\alpha\big)
 - (16 - b_{11}\epsilon - 4\epsilon^2) \big( Q_{S2} + F_{S2}^{(0)} \big) \,,
\\
\begin{split}
E_{S2}^{(2)} & = \bigg(\frac{\alpha}{4\pi}\bigg)^2
\big(\overline{s}_L^\alpha \gamma_{\mu_1} \gamma_{\mu_2}
                    \gamma_{\mu_3} \gamma_{\mu_4} \gamma_{\mu_5} d_L^\alpha\big) \otimes
\big(\overline{s}_L^\beta \gamma^{\mu_1} \gamma^{\mu_2}
     \gamma^{\mu_3} \gamma^{\mu_4} \gamma^{\mu_5} d_L^\beta\big)\\
 & \quad - \bigg(256 - a_{21}\epsilon - \frac{108816}{325}\epsilon^2\bigg) Q_{S2} \,,
\end{split}
\\
\begin{split}
F_{S2}^{(2)} & = \bigg(\frac{\alpha}{4\pi}\bigg)^2
\big(\overline{s}_L^\alpha \gamma_{\mu_1} \gamma_{\mu_2}
                    \gamma_{\mu_3} \gamma_{\mu_4} \gamma_{\mu_5} d_L^\beta\big) \otimes
\big(\overline{s}_L^\beta \gamma^{\mu_1} \gamma^{\mu_2}
     \gamma^{\mu_3} \gamma^{\mu_4} \gamma^{\mu_5} d_L^\alpha\big)\\
 & \quad - \bigg(256 - b_{21}\epsilon - \frac{108816}{325}\epsilon^2\bigg) \big( Q_{S2} + F_{S2}^{(0)} \big) \,.
\end{split}
\end{align}
The terms quadratic in $\epsilon$ do not play a role in our
calculation and are kept only for completeness. Some of the
coefficients of the $\epsilon$ terms have been left unspecified as an
additional check of our calculation. The result for the Wilson
coefficient at NLO in the electroweak interaction depends on the
coefficient $a_{11}$, as discussed in Sec.~\ref{sec:qcd}. Our results
given in this paper and all plots use the conventional choice $a_{11}
= 4$.

We expand the renormalization constants as
\begin{equation}
Z = 1 + \sum_{k=0}^\infty
\Bigg[ \sum_{i=1}^\infty Z^{(i,k)}
\bigg(\frac{\alpha_s}{4\pi}\bigg)^i
+ Z^{(e,k)} \frac{\alpha}{4\pi}
+ Z^{(se,k)} \frac{\alpha\alpha_s}{(4\pi)^2} \Bigg]
\frac{1}{\epsilon^k} \,.
\end{equation}
The necessary, non-zero $Z$ factors at order $\alpha_s$ are
\begin{equation}
Z_{Q_{S2}, Q_{S2}}^{(1,1)} = 2\,, \qquad
Z_{Q_{S2}, F_{S2}^{(0)}}^{(1,1)} = 3\,, \qquad
Z_{Q_{S2}, E_{S2}^{(1)}}^{(1,1)} = -\frac{1}{6}\,, \qquad
Z_{Q_{S2}, F_{S2}^{(1)}}^{(1,1)} = \frac{1}{2}\,,
\end{equation}
and
\begin{equation}
\begin{split}
Z_{F_{S2}^{(0)}, Q_{S2}}^{(1,0)} & = 2 - \frac{5}{12} a_{11} - \frac{1}{12} b_{11} \,, \qquad
Z_{E_{S2}^{(1)}, Q_{S2}}^{(1,0)} = - 48 - \frac{7}{3} a_{11} + 13 b_{11}
                                 + \frac{1}{6} a_{21} - \frac{1}{2} b_{21} \,, \\
Z_{F_{S2}^{(1)}, Q_{S2}}^{(1,0)} & = 48 + 5 a_{11} + \frac{65}{3} b_{11}
                                 - \frac{1}{4} a_{21} - \frac{7}{12} b_{21} \,.
\end{split}
\end{equation}
The necessary, non-zero $Z$ factors at order $\alpha$ are
\begin{equation}
Z_{Q_{S2}, Q_{S2}}^{(e,1)} = \frac{2}{3}\,, \qquad
Z_{Q_{S2}, E_{S2}^{(1)}}^{(e,1)} = \frac{1}{9}\,,
\end{equation}
and
\begin{equation}
Z_{F_{S2}^{(0)}, Q_{S2}}^{(e,0)} = \frac{1}{9} a_{11} - \frac{1}{9} b_{11} \,, \qquad
Z_{E_{S2}^{(1)}, Q_{S2}}^{(e,0)} = \frac{32}{9} a_{11} - \frac{1}{9} a_{21} \,, \qquad
Z_{F_{S2}^{(1)}, Q_{S2}}^{(e,0)} = \frac{32}{9} b_{11} - \frac{1}{9} b_{21} \,.
\end{equation}
At two-loop, we find
\begin{equation}
Z_{Q_{S2}, Q_{S2}}^{(se,1)} = - \frac{61}{9} - \frac{7}{27} a_{11} + \frac{13}{9} b_{11}
                       + \frac{1}{54} a_{21} - \frac{1}{18} b_{21} \,.
\end{equation}
Using these renormalization constants, we find that all scheme
dependence cancels in the two-loop anomalous dimension at order
$\alpha\alpha_s$, as expected on general grounds~\cite{Buras:1993dy}.

\section{Scheme independence at order $\alpha$}\label{sec:scheme}

In this appendix we give a proof of the scheme independence of the
prediction of the top-quark contribution to $\epsilon_K$ at NLO in
QED. We first recall the transformation properties of Wilson
coefficients and anomalous dimension at order $\alpha$ by adapting the
results in Ref.~\cite{Brod:2010mj}. To this end, we write the general
transformation among all dimension-six operators as
\begin{equation}\label{eq:gentmat}
\begin{pmatrix}
Q'\\
E'
\end{pmatrix}
=
\begin{pmatrix}
R&0\\
0&M
\end{pmatrix}
\begin{pmatrix}
1&0\\
\epsilon U + \epsilon^2 V&1
\end{pmatrix}
\begin{pmatrix}
1&W\\
0&1
\end{pmatrix}
\begin{pmatrix}
Q\\
E
\end{pmatrix}\,.
\end{equation}
Here, the matrices $R$ and $M$ parameterize a linear transformation
among the physical and evanescent operators $Q$ and $E$, respectively,
$W$ parameterizes the addition of multiples of evanescent operators to
the physical operators, and $U$ parameterizes the addition of
multiples of $\epsilon$ times physical operators to the evanescent
operators. As explained in detail in Ref.~\cite{Gorbahn:2004my}, this
transformation implies an additional finite renormalization that is
needed in order to restore the standard $\overline{\text{MS}}$
definition of the renormalization constants. This finite
renormalization, induced by the change~\eqref{eq:gentmat}, is given in
the notation of the previous section by
\begin{align}\label{eq:finz}
  Z_{QQ}^{\prime(1,0)} &= R \left[ W Z_{EQ}^{(1,0)} - \left( Z_{QE}^{(1,1)} +
      W Z_{EE}^{(1,1)} - \frac{1}{2} \gamma^{(0)} W \right) U \right]
  R^{-1}\,, \\ 
  Z_{QQ}^{\prime(e,0)} &= R \left[ W Z_{EQ}^{(e,0)} - \left( Z_{QE}^{(e,1)} +
      W Z_{EE}^{(e,1)} - \frac{1}{2} \gamma^{(e)} W \right) U \right]
  R^{-1}\,,
\end{align}
by a straightforward generalization of the results in
Ref.~\cite{Brod:2010mj}, where also the higher-order QCD expressions
can be found. The corresponding transformation law for the anomalous
dimension matrices is then
\begin{align}\label{eq:admtrafo}
\gamma^{\prime(0)} &= R\gamma^{(0)}R^{-1} \,,\\
\gamma^{\prime(e)} &= R\gamma^{(e)}R^{-1} \,,\\
\gamma^{\prime(se)} &= R\gamma^{(se)}R^{-1}
                - \left[Z_{QQ}^{\prime(e,0)},\gamma^{\prime(0)} \right]
                - \left[Z_{QQ}^{\prime(1,0)},\gamma^{\prime(e)} \right] \,,
\end{align}
in agreement with the results given in Ref.~\cite{Buras:1993dy}. The
Wilson coefficients change according to
\begin{equation}\label{eq:cprime}
  C' = \left [ 1 + \frac{\alpha_s }{4 \pi} Z^{\prime(1,0)}_{QQ}
                      + \frac{\alpha}{4 \pi} Z^{\prime(e,0)}_{QQ}
            \right ]^T \! \! ( R^{-1} \big )^T C \,.
\end{equation} 
In particular, we find
\begin{align}
  C^{\prime(e)}  & = C^{(e)} \,, \\
  C^{\prime(se)} & = C^{(se)} + C^{(0)} Z^{\prime(e,0)}_{QQ} + C^{(e)} Z^{\prime(1,0)}_{QQ} \,.
\end{align} 
The matrix elements change with the inverse transformation, thus
ensuring the the scheme dependence cancels in the
amplitude~\cite{Brod:2010mj}. Hence, we only need to show that the RG
evolution does not upend this cancelation. We will consider only the
effects of order $\alpha/\alpha_s$ and $\alpha$. (The QCD case has
been discussed extensively in the literature, see
e.g. Refs.~\cite{Buras:1991jm, Herrlich:1994kh, Gorbahn:2004my}. The
proof of scheme dependence including QED in Ref.~\cite{Buras:1993dy}
fails in our case as the anomalous dimension is itself scheme
independent.)

We expand the hadronic matrix element in powers of couplings
as\footnote{Of course, in reality this matrix element needs to be
  computed using nonperturbative methods such as lattice QCD. One can
  think of Eq.~\eqref{eq:expand:ME} as the perturbative conversion of
  the lattice result to the \MS{} scheme.}
\begin{equation}\label{eq:expand:ME}
  \langle Q \rangle (\mu)
= \bigg( 1 + \frac{\alpha_s(\mu)}{4\pi} r^{(1)}
           + \frac{\alpha}{4\pi} r^{(se)} \bigg)
  \langle Q \rangle^{(0)} \,.
\end{equation}
Recalling the perturbative expansion of the RG evolution matrix,
Eq.~\eqref{eq:U:expansion}, and the expansion in powers of couplings
of the low-energy Wilson coefficient,
Eq.~\eqref{eq:C:low:coefficients}, the terms at order
$\alpha/\alpha_s$ and $\alpha$ in the amplitude are then (we drop
obvious function arguments)
\begin{align}
  \langle CUQ \rangle^{(e)}
& = \frac{\alpha(\mu)}{\alpha_s(\mu)} \bigg( C^{(0)} U^{(e)} + \eta^{-1} C^{(e)} U^{(0)} \bigg)
    \langle Q \rangle^{(0)} \,, \\
\begin{split}
  \langle CUQ \rangle^{(se)}
& = \alpha(\mu) \bigg(   U^{(e)} \big[ \eta C^{(1)} + C^{(0)} r^{(1)} \big]
                       + \eta^{-1} C^{(e)} \big[ U^{(1)} + U^{(0)} r^{(1)} \big] \\
& \hspace{4em}         + C^{(0)} U^{(se)} + C^{(se)} U^{(0)} + C^{(0)} U^{(0)} r^{(se)}
                \bigg) \langle Q \rangle^{(0)} \,.
\end{split}
\end{align}
It is apparent that the order $\alpha/\alpha_s$ contribution is scheme
independent, as it involves only scheme-independent quantities. In the
order $\alpha$ contribution, the QCD scheme dependence cancels within
the square brackets in the first line, respectively, as shown in
Ref.~\cite{Buras:1991jm}. In the second line, note that the first term
is scheme independent. The renormalized matrix element $r^{(se)}$ is
given by\footnote{We assume that parameter and field renormalizations
  have been performed in the usual way.}
\begin{equation}
  r^{(se)} = r^{(se),\text{bare}} + Z^{(e)} \,.
\end{equation}
According to the discussion at the beginning of this section, under a
change of renormalization scheme with nonzero coefficient $U$ in
Eq.~\eqref{eq:gentmat}, this contribution to the matrix element
transforms as $r^{(se)} \to r^{(se)} - Z^{\prime (e)}$. This finishes
the proof of the scheme independence of the matrix element.

Note that for the cancelation to work, we need to fix the
electromagnetic coupling in the effective theory to its value at
$\mu=M_Z$, i.e. $\alpha = \alpha(M_Z)$. This is consistent with the
fixed-order perturbative expansion in $\alpha$.

\newpage

\section{The full two-loop result}\label{sec:full}

\begin{equation}
\begin{split}
C_{S2}^{tt,(se)} & = \big[128 y - 896 y z - 384 x y z + 288 x z^2 + 2752 y z^2 + 1504 x y z^2 + 448 x^2 y z^2 \\
      & \qquad - 1080 x^2 z^3 - 7536 x y z^3 + 872 x^2 y z^3 - 256 x^3 y z^3 - 576 y^2 z^3 + 144 x z^4 \\
      & \qquad - 864 x^3 z^4 + 608 y z^4 - 1240 x y z^4 + 1564 x^2 y z^4 - 1664 x^3 y z^4 + 64 x^4 y z^4 \\
      & \qquad + 2016 x y^2 z^4 + 288 y^3 z^4 - 540 x^2 z^5 + 6408 x^4 z^5 - 1632 x y z^5 + 4740 x^2 y z^5 \\
      & \qquad - 10694 x^3 y z^5 + 184 x^4 y z^5 - 3438 x^2 y^2 z^5 - 720 x y^3 z^5 + 432 x^3 z^6 \\
      & \qquad - 5184 x^5 z^6 + 2117 x^2 y z^6 - 1668 x^3 y z^6 - 4050 x^4 y z^6 + 1872 x^3 y^2 z^6 \\
      & \qquad + 765 x^2 y^3 z^6 - 36 x^4 z^7 + 432 x^6 z^7 - 373 x^3 y z^7 + 364 x^4 y z^7 \\
      & \qquad + 1080 x^5 y z^7 - 198 x^4 y^2 z^7 - 225 x^3 y^3 z^7\big]/\big[288 y (z-1) z (x z-1)^3\big]\\
& \quad + \pi^2 (-16 x + 128 x z + 16 x^2 z - 404 x z^2 - 24 x^2 z^2 + 8 x^3 z^2 + 472 x z^3 - 156 x^2 z^3 \\
      & \qquad \quad - 8 x^3 z^3 + 32 z^4 - 32 x z^4 + 148 x^2 z^4 - 134 x^3 z^4 + 48 x^4 z^4 + 8 x^5 z^4 - 16 z^5 \\
      & \qquad \quad + 64 x z^5 - 24 x^2 z^5 + 44 x^3 z^5 - 24 x^4 z^5 + 8 x^5 z^5 + 2 z^6 - 14 x z^6 + 2 x^2 z^6 \\
      & \qquad \quad - 96 x^4 z^6 + 2 x^5 z^6 - 18 x^2 y^2 z^6 - 6 x^2 z^7 - 108 x^5 z^7 - x^2 z^8)/\big[216 x^2 (z-1) z^3 \big]\\
& \quad + \Phi\bigg(\frac{1}{4x}\bigg) (-256 z + 2816 x z - 10624 x^2 z + 15232 x^3 z - 8192 x^4 z - 2048 x^5 z\\
      & \qquad \qquad \qquad + 128 z^2 - 512 x z^2 - 3584 x^2 z^2 + 21888 x^3 z^2 - 27520 x^4 z^2 + 11264 x^5 z^2\\
      & \qquad \qquad \qquad + 10240 x^6 z^2 - 16 z^3 - 32 x z^3 + 1304 x^2 z^3 - 4752 x^3 z^3 + 2928 x^4 z^3\\
      & \qquad \qquad \qquad - 12704 x^5 z^3 + 1024 x^6 z^3 - 14848 x^7 z^3 + 32 x z^4 - 1056 x^2 z^4 + 5936 x^3 z^4\\
      & \qquad \qquad \qquad - 4640 x^4 z^4 - 27952 x^5 z^4 + 72256 x^6 z^4 - 3840 x^7 z^4 + 7168 x^8 z^4\\
      & \qquad \qquad \qquad - 8 x^2 z^5 + 2304 x^3 z^5 - 19252 x^4 z^5 + 50320 x^5 z^5 - 35164 x^6 z^5\\
      & \qquad \qquad \qquad - 40352 x^7 z^5 + 19456 x^8 z^5 - 512 x^9 z^5 - 1532 x^4 z^6 + 11756 x^5 z^6\\
      & \qquad \qquad \qquad - 23860 x^6 z^6 + 4184 x^7 z^6 + 13696 x^8 z^6 - 2816 x^9 z^6 + 41 x^4 z^7\\
      & \qquad \qquad \qquad - 198 x^5 z^7 - 250 x^6 z^7 + 1612 x^7 z^7 - 252 x^8 z^7 - 1280 x^9 z^7)\\
      & \qquad \qquad \qquad /(576 x^2 (4x-1) (z-1) (x z-1)^4)\\
& \quad + \Phi\bigg(\frac{y}{4x}\bigg) (-176 x^2 z^3 + 256 x y z^3 - 56 y^2 z^3 + 656 x^3 z^4 - 1088 x^2 y z^4 + 304 x y^2 z^4\\
      & \qquad \qquad \qquad - 16 y^3 z^4 - 1164 x^4 z^5 + 1740 x^3 y z^5 - 564 x^2 y^2 z^5 + 48 x y^3 z^5 + 728 x^5 z^6\\
      & \qquad \qquad \qquad - 948 x^4 y z^6 + 324 x^3 y^2 z^6 - 32 x^2 y^3 z^6 - 44 x^6 z^7 + 112 x^5 y z^7\\
      & \qquad \qquad \qquad - 66 x^4 y^2 z^7 + 14 x^3 y^3 z^7 - 
    x^2 y^4 z^7) /(64 (z-1) (x z-1)^4)\\
& \quad + \Phi\bigg(\frac{z}{4}\bigg) (256 x^2 z + 7872 x^2 z^2 - 512 x^3 z^2 - 23072 x^2 z^3 - 17216 x^3 z^3 + 256 x^4 z^3\\
      & \qquad \qquad \qquad + 5664 x^2 z^4 - 1680 x^3 z^4 + 2048 x^4 z^4 + 5688 x^2 z^5 - 6240 x^3 z^5\\
      & \qquad \qquad \qquad - 2768 x^4 z^5 - 1222 x^2 z^6 - 1636 x^3 z^6 + 2032 x^4 z^6 - 73 x^2 z^7\\
      & \qquad \qquad \qquad + 900 x^3 z^7 + 322 x^4 z^7 + 32 x^3 z^8 - 208 x^4 z^8 - 8 x^4 z^9)\\
      & \qquad \qquad \qquad /(576 (z-1) (x z-1)^4)\\
\notag
\end{split}
\end{equation}
\begin{equation}
\begin{split}
& \quad + \Phi\bigg(\frac{yz}{4}\bigg) (144 x^2 z^3 + 48 x^3 z^4 - 224 x^2 y z^4 - 192 x^4 z^5 + 336 x^3 y z^5 + 
    132 x^2 y^2 z^5\\
      & \qquad \qquad \qquad - 88 x^4 y z^6 - 228 x^3 y^2 z^6 - 18 x^2 y^3 z^6 + 98 x^4 y^2 z^7 + 36 x^3 y^3 z^7\\
      & \qquad \qquad \qquad - x^2 y^4 z^7 - 16 x^4 y^3 z^8)/(64 (z-1) (x z-1)^4)\\
& \quad + \Phi\bigg(\frac{1}{xz}, \frac{1}{x}\bigg) (-64 + 576 z + 384 x z - 2128 z^2 - 2848 x z^2 - 928 x^2 z^2 + 3824 z^3\\
      & \qquad \qquad \qquad \quad + 8704 x z^3 + 5664 x^2 z^3 + 1120 x^3 z^3 - 3312 z^4 - 12752 x z^4\\
      & \qquad \qquad \qquad \quad - 14056 x^2 z^4 - 5816 x^3 z^4 - 672 x^4 z^4 + 992 z^5 + 9472 x z^5\\
      & \qquad \qquad \qquad \quad + 14904 x^2 z^5 + 11672 x^3 z^5 + 3544 x^4 z^5 + 224 x^5 z^5 + 272 z^6\\
      & \qquad \qquad \qquad \quad - 3032 x z^6 - 6904 x^2 z^6 - 9866 x^3 z^6 - 6800 x^4 z^6 - 1840 x^5 z^6\\
      & \qquad \qquad \qquad \quad - 224 x^6 z^6 - 144 z^7 - 424 x z^7 + 2232 x^2 z^7 + 4678 x^3 z^7\\
      & \qquad \qquad \qquad \quad + 6914 x^4 z^7 + 4536 x^5 z^7 + 1104 x^6 z^7 + 288 x^7 z^7 - 16 z^8\\
      & \qquad \qquad \qquad \quad + 456 x z^8 - 592 x^2 z^8 - 607 x^3 z^8 + 6444 x^4 z^8 + 368 x^5 z^8\\
      & \qquad \qquad \qquad \quad - 2664 x^6 z^8 - 408 x^7 z^8 - 160 x^8 z^8 + 40 x z^9 - 304 x^2 z^9\\
      & \qquad \qquad \qquad \quad - 1787 x^3 z^9 + 2891 x^4 z^9 + 2268 x^5 z^9 - 2720 x^6 z^9\\
      & \qquad \qquad \qquad \quad + 1000 x^7 z^9 - 8 x^8 z^9 + 32 x^9 z^9 - 16 x^2 z^{10} + 565 x^3 z^{10}\\
      & \qquad \qquad \qquad \quad - 1766 x^4 z^{10} + 2963 x^5 z^{10} - 2524 x^6 z^{10} + 1018 x^7 z^{10}\\
      & \qquad \qquad \qquad \quad - 272 x^8 z^{10} + 32 x^9 z^{10} + 41 x^3 z^{11} - 403 x^4 z^{11}\\
      & \qquad \qquad \qquad \quad + 1093 x^5 z^{11} - 1279 x^6 z^{11} + 694 x^7 z^{11} - 154 x^8 z^{11} + 8 x^9 z^{11})\\
      & \qquad \qquad \qquad \quad /(288 x^2 (z-1) z^4 (x z-1)^4)\\
& \quad +  \Phi\bigg(\frac{1}{xz}, \frac{y}{x}\bigg) (-16 + 104 x z + 64 y z - 296 x^2 z^2 - 296 x y z^2 - 96 y^2 z^2\\
      & \qquad \qquad \qquad \quad + 476 x^3 z^3 + 544 x^2 y z^3 + 344 x y^2 z^3 + 64 y^3 z^3 - 436 x^4 z^4\\
      & \qquad \qquad \qquad \quad - 568 x^3 y z^4 - 416 x^2 y^2 z^4 - 216 x y^3 z^4 - 16 y^4 z^4 + 160 x^5 z^5\\
      & \qquad \qquad \qquad \quad + 326 x^4 y z^5 + 271 x^3 y^2 z^5 + 192 x^2 y^3 z^5 + 64 x y^4 z^5 + 80 x^6 z^6\\
      & \qquad \qquad \qquad \quad - 106 x^5 y z^6 - 185 x^4 y^2 z^6 + 19 x^3 y^3 z^6 - 88 x^2 y^4 z^6 - 100 x^7 z^7\\
      & \qquad \qquad \qquad \quad + 114 x^6 y z^7 - 31 x^5 y^2 z^7 - 24 x^4 y^3 z^7 + 41 x^3 y^4 z^7 + 28 x^8 z^8\\
      & \qquad \qquad \qquad \quad - 78 x^7 y z^8 + 65 x^6 y^2 z^8 - 7 x^5 y^3 z^8 - 9 x^4 y^4 z^8 + x^3 y^5 z^8)\\
      & \qquad \qquad \qquad \quad /(32 x^2 (z-1) z (x z-1)^4)\\
& \quad + \text{Li}_2 \bigg(1 - \frac{1}{z}\bigg) (16 - 112 z - 48 x z + 292 z^2 + 232 x z^2 + 40 x^2 z^2 - 260 z^3\\
      & \qquad \qquad \qquad \quad - 420 x z^3 - 128 x^2 z^3 + 16 z^4 + 236 x z^4 + 114 x^2 z^4 - 40 x^3 z^4\\
      & \qquad \qquad \qquad \quad + 44 z^5 - 64 x z^5 + 186 x^2 z^5 + 72 x^3 z^5 + 4 z^6 - 42 x z^6 + 8 x^2 z^6\\
      & \qquad \qquad \qquad \quad - 155 x^3 z^6 - 2 x z^7 - 54 x^2 z^7 + 31 x^3 z^7 - 4 x^2 z^8 + 36 x^3 z^8 + 2 x^3 z^9)\\
      & \qquad \qquad \qquad \quad/(36 x z^3 (x z-1)^2)\\
\notag
\end{split}
\end{equation}
\begin{equation}
\begin{split}
& \quad + \text{Li}_2 \big(1 - yz \big) (4 - 14 x z - 12 y z + 20 x^2 z^2 + 24 x y z^2 + 12 y^2 z^2 - 13 x^3 z^3\\
      & \qquad \qquad \qquad \qquad - 12 x^2 y z^3 - 18 x y^2 z^3 - 4 y^3 z^3 + 6 x^3 y z^4 - 6 x^2 y^2 z^4 + 8 x y^3 z^4\\
      & \qquad \qquad \qquad \qquad + 9 x^3 y^2 z^5 - 2 x^2 y^3 z^5 - 2 x^3 y^3 z^6)/(4 x (z-1) (x z-1)^2)\\
& \quad + \text{Li}_2 \big(1 - xz \big) (-32 + 256 z + 128 x z - 808 z^2 - 816 x z^2 - 176 x^2 z^2 + 944 z^3\\
      & \qquad \qquad \qquad \qquad + 2112 x z^3 + 896 x^2 z^3 + 80 x^3 z^3 - 2504 x z^4 - 1788 x^2 z^4\\
      & \qquad \qquad \qquad \qquad - 396 x^3 z^4 + 1968 x^2 z^5 + 616 x^3 z^5 + 164 x^4 z^5 + 32 x^5 z^5\\
      & \qquad \qquad \qquad \qquad - 3 x^3 z^6 - 644 x^4 z^6 - 164 x^5 z^6 - 48 x^6 z^6 - 375 x^4 z^7 + 332 x^5 z^7\\
      & \qquad \qquad \qquad \qquad + 44 x^6 z^7 + 16 x^7 z^7 + 25 x^5 z^8 - 112 x^6 z^8 + 16 x^7 z^8 + 449 x^6 z^9\\
      & \qquad \qquad \qquad \qquad + 4 x^7 z^9 - 216 x^7 z^{10})/(72 x (z-1) z^3 (x z-1)^3)\\
& \quad + \text{Li}_2 \bigg(1 - \frac{x}{y} \bigg) (9 x^2 z^3 - 18 x y z^3 + 9 y^2 z^3 - 20 x^3 z^4 + 42 x^2 y z^4 - 24 x y^2 z^4\\
      & \qquad \qquad \qquad \qquad + 2 y^3 z^4 + 8 x^4 z^5 - 18 x^3 y z^5 + 12 x^2 y^2 z^5 - 2 x y^3 z^5)\\
      & \qquad \qquad \qquad \qquad /(4 (z-1) (x z-1)^2)\\
& \quad + \text{Li}_2 \big(1 - x \big) (-64 z + 192 x z - 160 x^2 z + 16 x^3 z + 16 x^5 z + 32 z^2 - 32 x z^2\\
      & \qquad \qquad \qquad \qquad - 72 x^2 z^2 - 16 x^3 z^2 + 200 x^4 z^2 - 96 x^5 z^2 - 16 x^6 z^2 - 4 z^3\\
      & \qquad \qquad \qquad \qquad - 2 x z^3 + 141 x^2 z^3 - 256 x^3 z^3 + 165 x^4 z^3 - 84 x^5 z^3 + 32 x^6 z^3\\
      & \qquad \qquad \qquad \qquad + 8 x^7 z^3 + 4 x z^4 - 98 x^2 z^4 + 136 x^3 z^4 - 6 x^4 z^4 - 16 x^5 z^4\\
      & \qquad \qquad \qquad \qquad - 28 x^6 z^4 + 8 x^7 z^4 + 84 x^3 z^5 - 188 x^4 z^5 + 126 x^5 z^5\\
      & \qquad \qquad \qquad \qquad - 24 x^6 z^5 + 2 x^7 z^5)/(36 x^2 (z-1) (x z-1)^2)\\
& \quad + \log^2(x) (256 y z - 1280 x y z + 1408 x^2 y z - 128 x^3 y z - 128 x^5 y z - 128 y z^2\\
      & \qquad \qquad \qquad - 512 x y z^2 + 4096 x^2 y z^2 - 4352 x^3 y z^2 - 1216 x^4 y z^2 + 768 x^5 y z^2\\
      & \qquad \qquad \qquad + 512 x^6 y z^2 + 16 y z^3 + 256 x y z^3 - 376 x^2 y z^3 - 2688 x^3 y z^3\\
      & \qquad \qquad \qquad + 3776 x^4 y z^3 + 8160 x^5 y z^3 - 2560 x^6 y z^3 - 832 x^7 y z^3 + 1296 x^3 y^2 z^3\\
      & \qquad \qquad \qquad - 504 x^2 y^3 z^3 - 48 x y z^4 + 736 x^2 y z^4 + 72 x^3 y z^4 - 3904 x^4 y z^4\\
      & \qquad \qquad \qquad - 7296 x^5 y z^4 - 12608 x^6 y z^4 + 3008 x^7 y z^4 + 704 x^8 y z^4 - 6480 x^4 y^2 z^4\\
      & \qquad \qquad \qquad + 2952 x^3 y^3 z^4 - 144 x^2 y^4 z^4 - 5184 x^7 z^5 + 40 x^2 y z^5 - 3120 x^3 y z^5\\
      & \qquad \qquad \qquad + 6228 x^4 y z^5 + 968 x^5 y z^5 + 13272 x^6 y z^5 + 6000 x^7 y z^5 - 1344 x^8 y z^5\\
      & \qquad \qquad \qquad - 320 x^9 y z^5 + 13284 x^5 y^2 z^5 - 6660 x^4 y^3 z^5 + 576 x^3 y^4 z^5\\
      & \qquad \qquad \qquad + 10368 x^8 z^6 - 8 x^3 y z^6 + 3788 x^4 y z^6 - 8232 x^5 y z^6 + 2092 x^6 y z^6\\
      & \qquad \qquad \qquad - 2760 x^7 y z^6 + 48 x^8 y z^6 + 64 x^9 y z^6 + 64 x^{10} y z^6 - 12528 x^6 y^2 z^6\\
      & \qquad \qquad \qquad + 6552 x^5 y^3 z^6 - 720 x^4 y^4 z^6 - 5184 x^9 z^7 - 41 x^4 y z^7 - 1580 x^5 y z^7\\
      & \qquad \qquad \qquad + 2772 x^6 y z^7 + 804 x^7 y z^7 - 5584 x^8 y z^7 - 272 x^9 y z^7 + 64 x^{10} y z^7\\
      & \qquad \qquad \qquad + 4824 x^7 y^2 z^7 - 2736 x^6 y^3 z^7 + 396 x^5 y^4 z^7 - 9 x^4 y^5 z^7 + 41 x^5 y z^8\\
\notag
\end{split}
\end{equation}
\begin{equation}
\begin{split}
      & \qquad \qquad \qquad + 48 x^6 y z^8 - 208 x^7 y z^8 - 88 x^8 y z^8 + 3696 x^9 y z^8 + 16 x^{10} y z^8\\
      & \qquad \qquad \qquad - 396 x^8 y^2 z^8 + 396 x^7 y^3 z^8 - 108 x^6 y^4 z^8 + 9 x^5 y^5 z^8 - 864 x^{10} y z^9)\\
      & \qquad \qquad \qquad /(576 x^2 y (z-1) (x z-1)^5)\\
& \quad + \log^2(y) (9 x^2 z^3 - 18 x y z^3 + 7 y^2 z^3 - 20 x^3 z^4 + 42 x^2 y z^4 - 20 x y^2 z^4 + 2 y^3 z^4\\
      & \qquad \qquad \qquad + 8 x^4 z^5 - 18 x^3 y z^5 + 10 x^2 y^2 z^5 - 2 x y^3 z^5)/(8 (z-1) (x z-1)^2)\\
& \quad + \log(z) (128 y - 832 y z - 448 x y z + 1776 y z^2 + 2976 x y z^2 + 512 x^2 y z^2 - 192 y z^3\\
      & \qquad \qquad \qquad - 6256 x y z^3 - 4784 x^2 y z^3 - 160 x^3 y z^3 - 144 y^2 z^3 + 216 x z^4\\
      & \qquad \qquad \qquad - 648 x^3 z^4 - 16 y z^4 + 232 x y z^4 + 10336 x^2 y z^4 + 4032 x^3 y z^4\\
      & \qquad \qquad \qquad - 64 x^4 y z^4 + 684 x y^2 z^4 - 1026 x^2 z^5 + 3888 x^4 z^5 + 40 x y z^5\\
      & \qquad \qquad \qquad + 268 x^2 y z^5 - 5374 x^3 y z^5 - 1576 x^4 y z^5 + 32 x^5 y z^5 - 1134 x^2 y^2 z^5\\
      & \qquad \qquad \qquad + 1566 x^3 z^6 - 7128 x^5 z^6 - 491 x^2 y z^6 - 1668 x^3 y z^6 + 9007 x^4 y z^6\\
      & \qquad \qquad \qquad + 184 x^5 y z^6 + 1377 x^3 y^2 z^6 - 171 x^2 y^3 z^6 - 810 x^4 z^7 + 4104 x^6 z^7\\
      & \qquad \qquad \qquad + 203 x^3 y z^7 + 340 x^4 y z^7 - 1938 x^5 y z^7 - 648 x^4 y^2 z^7 + 189 x^3 y^3 z^7\\
      & \qquad \qquad \qquad + 54 x^5 z^8 - 216 x^7 z^8 - 96 x^4 y z^8 - 78 x^5 y z^8 + 567 x^6 y z^8\\
      & \qquad \qquad \qquad + 27 x^5 y^2 z^8 - 72 x^4 y^3 z^8)/(144 y (z-1) z (x z-1)^4)\\
& \quad + \log^2(z) (-288 y + 1872 x y z - 560 x^3 y z + 864 y^2 z - 5328 x^2 y z^2 + 1424 x^3 y z^2\\
      & \qquad \qquad \qquad + 2096 x^4 y z^2 + 128 x^5 y z^2 - 4320 x y^2 z^2 - 864 y^3 z^2 + 2348 x^3 y z^3\\
      & \qquad \qquad \qquad - 2672 x^4 y z^3 - 3168 x^5 y z^3 - 448 x^6 y z^3 + 8640 x^2 y^2 z^3 + 3888 x y^3 z^3\\
      & \qquad \qquad \qquad + 288 y^4 z^3 + 8248 x^3 y z^4 - 5916 x^4 y z^4 + 2400 x^5 y z^4 + 2464 x^6 y z^4\\
      & \qquad \qquad \qquad + 576 x^7 y z^4 - 9864 x^3 y^2 z^4 - 6048 x^2 y^3 z^4 - 1440 x y^4 z^4 + 1296 x^4 z^5\\
      & \qquad \qquad \qquad - 5184 x^6 z^5 + 2806 x^3 y z^5 - 16040 x^4 y z^5 + 20288 x^5 y z^5 - 2928 x^6 y z^5\\
      & \qquad \qquad \qquad - 944 x^7 y z^5 - 320 x^8 y z^5 + 7920 x^4 y^2 z^5 + 3618 x^3 y^3 z^5 + 2736 x^2 y^4 z^5\\
      & \qquad \qquad \qquad - 2592 x^5 z^6 + 10368 x^7 z^6 - 1368 x^3 y z^6 - 2450 x^4 y z^6 + 9708 x^5 y z^6\\
      & \qquad \qquad \qquad - 11280 x^6 y z^6 + 2240 x^7 y z^6 + 48 x^8 y z^6 + 64 x^9 y z^6 - 4212 x^5 y^2 z^6\\
      & \qquad \qquad \qquad - 270 x^4 y^3 z^6 - 2484 x^3 y^4 z^6 + 1296 x^6 z^7 - 5184 x^8 z^7 - 73 x^3 y z^7\\
      & \qquad \qquad \qquad + 2332 x^4 y z^7 - 498 x^5 y z^7 - 1916 x^6 y z^7 - 3060 x^7 y z^7 - 480 x^8 y z^7\\
      & \qquad \qquad \qquad + 64 x^9 y z^7 + 972 x^6 y^2 z^7 - 378 x^5 y^3 z^7 + 1080 x^4 y^4 z^7 - 9 x^3 y^5 z^7\\
      & \qquad \qquad \qquad + 105 x^4 y z^8 - 1188 x^5 y z^8 + 142 x^6 y z^8 + 3524 x^8 y z^8 + 16 x^9 y z^8\\
      & \qquad \qquad \qquad + 54 x^6 y^3 z^8 - 180 x^5 y^4 z^8 + 9 x^4 y^5 z^8 - 40 x^5 y z^9 + 224 x^6 y z^9\\
      & \qquad \qquad \qquad - 864 x^9 y z^9 + 8 x^6 y z^{10})/(576 x y (z-1) (x z-1)^5)\\
& \quad + \log\bigg(\frac{\mu}{M_W}\bigg) ((432 x z - 1408 x y z - 1620 x^2 z^2 + 1072 x y z^2 + 5280 x^2 y z^2\\
      & \qquad \qquad \qquad \quad + 216 x z^3 + 432 x^3 z^3 - 312 x y z^3 - 3732 x^2 y z^3 - 4224 x^3 y z^3\\
      & \qquad \qquad \qquad \quad + 108 x y^2 z^3 - 810 x^2 z^4 + 3132 x^4 z^4 + 1323 x^2 y z^4 - 312 x^3 y z^4\\
      & \qquad \qquad \qquad \quad + 352 x^4 y z^4 - 405 x^2 y^2 z^4 + 648 x^3 z^5 - 2592 x^5 z^5 - 171 x^3 y z^5\\
      & \qquad \qquad \qquad \quad + 137 x^4 y z^5 + 324 x^3 y^2 z^5 - 54 x^4 z^6 + 216 x^6 z^6 + 78 x^4 y z^6\\
      & \qquad \qquad \qquad \quad - 135 x^5 y z^6 - 27 x^4 y^2 z^6)/(72 y (z-1) (x z-1)^3)\\
\notag
\end{split}
\end{equation}
\begin{equation}
\begin{split}
      & \qquad \qquad \qquad \quad + (108 x^3 z^3 - 352 x^3 y z^3 - 108 x^4 z^4 + 436 x^3 y z^4 + 352 x^4 y z^4\\
      & \qquad \qquad \qquad \quad + 54 x^3 z^5 - 216 x^5 z^5 - 180 x^3 y z^5 + 113 x^4 y z^5 + 27 x^3 y^2 z^5\\
      & \qquad \qquad \qquad \quad - 54 x^4 z^6 + 216 x^6 z^6 + 27 x^4 y z^6 - 54 x^5 y z^6 - 27 x^4 y^2 z^6)\\
      & \qquad \qquad \qquad \quad \times \log(xz)/(12 y (z-1) (x z-1)^4)\\
& \quad + \log(y) ((-16 y z^2 + 40 x y z^3 + 16 y^2 z^3 - 30 x^2 y z^4 - 40 x y^2 z^4 + 7 x^3 y z^5\\
      & \qquad \qquad \qquad + 43 x^2 y^2 z^5 - x^4 y z^6 - 13 x^3 y^2 z^6)/(16 (z-1) (x z-1)^3)\\
      & \qquad \qquad \qquad + (16 - 88 x z - 48 y z + 208 x^2 z^2 + 192 x y z^2 + 48 y^2 z^2 - 268 x^3 z^3\\
      & \qquad \qquad \qquad - 288 x^2 y z^3 - 168 x y^2 z^3 - 16 y^3 z^3 + 168 x^4 z^4 + 284 x^3 y z^4\\
      & \qquad \qquad \qquad + 168 x^2 y^2 z^4 + 64 x y^3 z^4 + 8 x^5 z^5 - 210 x^4 y z^5 - 39 x^3 y^2 z^5\\
      & \qquad \qquad \qquad - 88 x^2 y^3 z^5 - 72 x^6 z^6 + 120 x^5 y z^6 - 48 x^4 y^2 z^6 + 58 x^3 y^3 z^6\\
      & \qquad \qquad \qquad + 28 x^7 z^7 - 50 x^6 y z^7 + 21 x^5 y^2 z^7 - 12 x^4 y^3 z^7)\\
      & \qquad \qquad \qquad \times \log(z)/(32 x (z-1) (x z-1)^4))\\
& \quad + \log(x) ((-64 y + 256 x y + 448 y z - 1600 x y z - 768 x^2 y z - 1248 y z^2 + 2528 x y z^2\\
      & \qquad \qquad \qquad + 6688 x^2 y z^2 + 384 x^3 y z^2 - 64 y z^3 + 5360 x y z^3 - 10944 x^2 y z^3\\
      & \qquad \qquad \qquad - 13504 x^3 y z^3 + 768 x^4 y z^3 + 288 y^2 z^3 - 1152 x y^2 z^3 + 648 x^3 z^4\\
      & \qquad \qquad \qquad - 2592 x^4 z^4 - 296 y z^4 + 1496 x y z^4 - 11792 x^2 y z^4 + 19904 x^3 y z^4\\
      & \qquad \qquad \qquad + 14048 x^4 y z^4 - 896 x^5 y z^4 - 1188 x y^2 z^4 + 4752 x^2 y^2 z^4 - 144 y^3 z^4\\
      & \qquad \qquad \qquad + 576 x y^3 z^4 - 3888 x^4 z^5 + 15552 x^5 z^5 + 1104 x y z^5 - 4900 x^2 y z^5\\
      & \qquad \qquad \qquad + 11174 x^3 y z^5 - 7328 x^4 y z^5 - 6432 x^5 y z^5 + 256 x^6 y z^5 + 1764 x^2 y^2 z^5\\
      & \qquad \qquad \qquad - 7056 x^3 y^2 z^5 + 504 x y^3 z^5 - 2016 x^2 y^3 z^5 - 108 x^3 z^6 + 432 x^4 z^6\\
      & \qquad \qquad \qquad + 7128 x^5 z^6 - 28512 x^6 z^6 - 1432 x^2 y z^6 + 7409 x^3 y z^6 - 20447 x^4 y z^6\\
      & \qquad \qquad \qquad + 33124 x^5 y z^6 + 864 x^6 y z^6 - 1710 x^3 y^2 z^6 + 6840 x^4 y^2 z^6 - 576 x^2 y^3 z^6\\
      & \qquad \qquad \qquad + 2304 x^3 y^3 z^6 + 108 x^4 z^7 - 432 x^5 z^7 - 4104 x^6 z^7 + 16416 x^7 z^7\\
      & \qquad \qquad \qquad + 1099 x^3 y z^7 - 4722 x^4 y z^7 + 5226 x^5 y z^7 - 7528 x^6 y z^7 + 720 x^4 y^2 z^7\\
      & \qquad \qquad \qquad - 2880 x^5 y^2 z^7 + 315 x^3 y^3 z^7 - 1260 x^4 y^3 z^7 + 216 x^7 z^8 - 864 x^8 z^8\\
      & \qquad \qquad \qquad - 115 x^4 y z^8 + 439 x^5 y z^8 - 783 x^6 y z^8 + 2268 x^7 y z^8 - 36 x^5 y^2 z^8\\
      & \qquad \qquad \qquad + 144 x^6 y^2 z^8 - 45 x^4 y^3 z^8 + 180 x^5 y^3 z^8)\\
      & \qquad \qquad \qquad /(144 (4 x-1) y (z-1) z (x z-1)^4)\\
      & \qquad \qquad \qquad + (16 - 88 x z - 48 y z + 208 x^2 z^2 + 192 x y z^2 + 48 y^2 z^2 - 340 x^3 z^3\\
      & \qquad \qquad \qquad - 144 x^2 y z^3 - 224 x y^2 z^3 - 16 y^3 z^3 + 472 x^4 z^4 - 340 x^3 y z^4\\
      & \qquad \qquad \qquad + 440 x^2 y^2 z^4 + 48 x y^3 z^4 - 448 x^5 z^5 + 750 x^4 y z^5 - 495 x^3 y^2 z^5\\
      & \qquad \qquad \qquad - 40 x^2 y^3 z^5 + 216 x^6 z^6 - 504 x^5 y z^6 + 272 x^4 y^2 z^6 + 10 x^3 y^3 z^6\\
      & \qquad \qquad \qquad - 36 x^7 z^7 + 94 x^6 y z^7 - 59 x^5 y^2 z^7 + 4 x^4 y^3 z^7)\\
      & \qquad \qquad \qquad \times \log(y)/(32 x (z-1) (x z-1)^4)\\
\notag
\end{split}
\end{equation}
\begin{equation}
\begin{split}
      & \qquad \qquad \qquad + ((-64 y + 512 y z + 384 x y z - 1616 y z^2 - 2656 x y z^2 - 928 x^2 y z^2\\
      & \qquad \qquad \qquad + 2064 y z^3 + 7456 x y z^3 + 5568 x^2 y z^3 + 1120 x^3 y z^3 - 1104 y z^4\\
      & \qquad \qquad \qquad - 8312 x y z^4 - 13640 x^2 y z^4 - 6008 x^3 y z^4 - 672 x^4 y z^4 + 432 y^2 z^4\\
      & \qquad \qquad \qquad - 112 y z^5 + 4512 x y z^5 + 11544 x^2 y z^5 + 12608 x^3 y z^5 + 3704 x^4 y z^5\\
      & \qquad \qquad \qquad + 224 x^5 y z^5 - 2160 x y^2 z^5 - 432 y^3 z^5 + 160 y z^6 + 248 x y z^6\\
      & \qquad \qquad \qquad - 6200 x^2 y z^6 - 6630 x^3 y z^6 - 3968 x^4 y z^6 - 1776 x^5 y z^6 - 224 x^6 y z^6\\
      & \qquad \qquad \qquad + 4320 x^2 y^2 z^6 + 1944 x y^3 z^6 + 144 y^4 z^6 + 16 y z^7 - 640 x y z^7\\
      & \qquad \qquad \qquad + 176 x^2 y z^7 + 5732 x^3 y z^7 - 10218 x^4 y z^7 - 2248 x^5 y z^7 + 1072 x^6 y z^7\\
      & \qquad \qquad \qquad + 288 x^7 y z^7 - 4716 x^3 y^2 z^7 - 3024 x^2 y^3 z^7 - 720 x y^4 z^7 + 648 x^4 z^8\\
      & \qquad \qquad \qquad - 5184 x^6 z^8 - 56 x y z^8 + 760 x^2 y z^8 + 1197 x^3 y z^8 - 6256 x^4 y z^8\\
      & \qquad \qquad \qquad + 25032 x^5 y z^8 + 520 x^6 y z^8 - 472 x^7 y z^8 - 160 x^8 y z^8 + 3474 x^4 y^2 z^8\\
      & \qquad \qquad \qquad + 1755 x^3 y^3 z^8 + 1368 x^2 y^4 z^8 - 1296 x^5 z^9 + 10368 x^7 z^9 + 56 x^2 y z^9\\
      & \qquad \qquad \qquad - 870 x^3 y z^9 - 53 x^4 y z^9 + 2044 x^5 y z^9 - 11776 x^6 y z^9 + 1120 x^7 y z^9\\
      & \qquad \qquad \qquad + 24 x^8 y z^9 + 32 x^9 y z^9 - 1242 x^5 y^2 z^9 - 351 x^4 y^3 z^9 - 1170 x^3 y^4 z^9\\
      & \qquad \qquad \qquad + 648 x^6 z^{10} - 5184 x^8 z^{10} - 57 x^3 y z^{10} + 952 x^4 y z^{10} - 2187 x^5 y z^{10}\\
      & \qquad \qquad \qquad + 1820 x^6 y z^{10} - 2962 x^7 y z^{10} - 240 x^8 y z^{10} + 32 x^9 y z^{10}\\
      & \qquad \qquad \qquad - 558 x^6 y^2 z^{10} + 243 x^5 y^3 z^{10} + 450 x^4 y^4 z^{10} - 9 x^3 y^5 z^{10}\\
      & \qquad \qquad \qquad + 41 x^4 y z^{11} - 362 x^5 y z^{11} + 731 x^6 y z^{11} - 548 x^7 y z^{11}\\
      & \qquad \qquad \qquad + 3418 x^8 y z^{11} + 8 x^9 y z^{11} + 450 x^7 y^2 z^{11} - 135 x^6 y^3 z^{11}\\
      & \qquad \qquad \qquad - 72 x^5 y^4 z^{11} + 9 x^4 y^5 z^{11} - 864 x^9 y z^{12})\\
      & \qquad \qquad \qquad \times \log(z))/(288 x y (z-1) z^3 (x z-1)^5) )\,.
\end{split}
\end{equation}
Here, $x \equiv m_t^2/M_Z^2$, $y \equiv M_h^2/M_Z^2$, and $z \equiv
M_Z^2/M_w^2$. The functions $\Phi$ are defined in
Ref.~\cite{Davydychev:1992mt}.

\addcontentsline{toc}{section}{References}
\bibliographystyle{JHEP}
\bibliography{references}

\providecommand{\href}[2]{#2}\begingroup\raggedright\begin{thebibliography}{10}

\bibitem{Proceedings:2001rdi}
\emph{{$B$ physics at the Tevatron: Run II and beyond}}, 12, 2001.

\bibitem{ParticleDataGroup:2020ssz}
{\scshape Particle Data Group} collaboration, P.~A. Zyla et~al., \emph{{Review
  of Particle Physics}},
  \href{https://doi.org/10.1093/ptep/ptaa104}{\emph{PTEP} {\bfseries 2020}
  (2020) 083C01}.

\bibitem{Aoki:2019cca}
{\scshape Flavour Lattice Averaging Group} collaboration, S.~Aoki et~al.,
  \emph{{FLAG Review 2019}},
  \href{https://arxiv.org/abs/1902.08191}{{\ttfamily 1902.08191}}.

\bibitem{Buras:2010pza}
A.~J. Buras, D.~Guadagnoli and G.~Isidori, \emph{{On $\epsilon_K$ Beyond Lowest
  Order in the Operator Product Expansion}},
  \href{https://doi.org/10.1016/j.physletb.2010.04.017}{\emph{Phys. Lett.}
  {\bfseries B688} (2010) 309--313},
  [\href{https://arxiv.org/abs/1002.3612}{{\ttfamily 1002.3612}}].

\bibitem{Inami:1980fz}
T.~Inami and C.~S. Lim, \emph{{Effects of Superheavy Quarks and Leptons in
  Low-Energy Weak Processes $K_L \to \mu \bar \mu$, $K^+ \to \pi^+ \nu \bar
  \nu$, and $K^0 \leftrightarrow \bar K^0$}},
  \href{https://doi.org/10.1143/PTP.65.297}{\emph{Prog. Theor. Phys.}
  {\bfseries 65} (1981) 297}.

\bibitem{Buchalla:1995vs}
G.~Buchalla, A.~J. Buras and M.~E. Lautenbacher, \emph{{Weak decays beyond
  leading logarithms}},
  \href{https://doi.org/10.1103/RevModPhys.68.1125}{\emph{Rev. Mod. Phys.}
  {\bfseries 68} (1996) 1125--1144},
  [\href{https://arxiv.org/abs/hep-ph/9512380}{{\ttfamily hep-ph/9512380}}].

\bibitem{Brod:2019rzc}
J.~Brod, M.~Gorbahn and E.~Stamou, \emph{{Standard-Model Prediction of
  $\epsilon_K$ with Manifest Quark-Mixing Unitarity}},
  \href{https://doi.org/10.1103/PhysRevLett.125.171803}{\emph{Phys. Rev. Lett.}
  {\bfseries 125} (2020) 171803},
  [\href{https://arxiv.org/abs/1911.06822}{{\ttfamily 1911.06822}}].

\bibitem{Buras:1990fn}
A.~J. Buras, M.~Jamin and P.~H. Weisz, \emph{{Leading and Next-to-leading
  \{QCD\} Corrections to $\epsilon$ Parameter and $B^0 - \bar{B}^0$ Mixing in
  the Presence of a Heavy Top Quark}},
  \href{https://doi.org/10.1016/0550-3213(90)90373-L}{\emph{Nucl. Phys. B}
  {\bfseries 347} (1990) 491--536}.

\bibitem{Herrlich:1993yv}
S.~Herrlich and U.~Nierste, \emph{{Enhancement of the K(L) - K(S) mass
  difference by short distance QCD corrections beyond leading logarithms}},
  \href{https://doi.org/10.1016/0550-3213(94)90044-2}{\emph{Nucl. Phys.}
  {\bfseries B419} (1994) 292--322},
  [\href{https://arxiv.org/abs/hep-ph/9310311}{{\ttfamily hep-ph/9310311}}].

\bibitem{Brod:2010mj}
J.~Brod and M.~Gorbahn, \emph{{$\epsilon_K$ at Next-to-Next-to-Leading Order:
  The Charm-Top-Quark Contribution}},
  \href{https://doi.org/10.1103/PhysRevD.82.094026}{\emph{Phys. Rev. D}
  {\bfseries 82} (2010) 094026},
  [\href{https://arxiv.org/abs/1007.0684}{{\ttfamily 1007.0684}}].

\bibitem{Brod:2011ty}
J.~Brod and M.~Gorbahn, \emph{{Next-to-Next-to-Leading-Order Charm-Quark
  Contribution to the CP Violation Parameter $\epsilon_K$ and $\Delta M_K$}},
  \href{https://doi.org/10.1103/PhysRevLett.108.121801}{\emph{Phys. Rev. Lett.}
  {\bfseries 108} (2012) 121801},
  [\href{https://arxiv.org/abs/1108.2036}{{\ttfamily 1108.2036}}].

\bibitem{BGSY}
J.~Brod, M.~Gorbahn, E.~Stamou and H.~Yu, \emph{{work in progress}}, .

\bibitem{Gambino:1998rt}
P.~Gambino, A.~Kwiatkowski and N.~Pott, \emph{{Electroweak effects in the B0 -
  anti-B0 mixing}},
  \href{https://doi.org/10.1016/S0550-3213(98)00860-8}{\emph{Nucl. Phys. B}
  {\bfseries 544} (1999) 532--556},
  [\href{https://arxiv.org/abs/hep-ph/9810400}{{\ttfamily hep-ph/9810400}}].

\bibitem{Bobeth:2013tba}
C.~Bobeth, M.~Gorbahn and E.~Stamou, \emph{{Electroweak Corrections to $B_{s,d}
  \to \ell^+ \ell^-$}},
  \href{https://doi.org/10.1103/PhysRevD.89.034023}{\emph{Phys. Rev. D}
  {\bfseries 89} (2014) 034023},
  [\href{https://arxiv.org/abs/1311.1348}{{\ttfamily 1311.1348}}].

\bibitem{Buras:2002vd}
A.~J. Buras, P.~H. Chankowski, J.~Rosiek and L.~Slawianowska, \emph{{$\Delta
  M_{d,s}, B^0{d,s} \to \mu^{+} \mu^{-}$ and $B \to X_{s} \gamma$ in
  supersymmetry at large $\tan\beta$}},
  \href{https://doi.org/10.1016/S0550-3213(03)00190-1}{\emph{Nucl. Phys. B}
  {\bfseries 659} (2003) 3},
  [\href{https://arxiv.org/abs/hep-ph/0210145}{{\ttfamily hep-ph/0210145}}].

\bibitem{Vermaseren:2000nd}
J.~A.~M. Vermaseren, \emph{{New features of FORM}},
  \href{https://arxiv.org/abs/math-ph/0010025}{{\ttfamily math-ph/0010025}}.

\bibitem{Davydychev:1992mt}
A.~I. Davydychev and J.~B. Tausk, \emph{{Two loop selfenergy diagrams with
  different masses and the momentum expansion}},
  \href{https://doi.org/10.1016/0550-3213(93)90338-P}{\emph{Nucl. Phys.}
  {\bfseries B397} (1993) 123--142}.

\bibitem{Bobeth:1999mk}
C.~Bobeth, M.~Misiak and J.~Urban, \emph{{Photonic penguins at two loops and
  $m_t$ dependence of $BR[B \to X_s l^+ l^-]$}},
  \href{https://doi.org/10.1016/S0550-3213(00)00007-9}{\emph{Nucl. Phys.}
  {\bfseries B574} (2000) 291--330},
  [\href{https://arxiv.org/abs/hep-ph/9910220}{{\ttfamily hep-ph/9910220}}].

\bibitem{Nogueira:1991ex}
P.~Nogueira, \emph{{Automatic Feynman graph generation}},
  \href{https://doi.org/10.1006/jcph.1993.1074}{\emph{J. Comput. Phys.}
  {\bfseries 105} (1993) 279--289}.

\bibitem{Sirlin:1981ie}
A.~Sirlin, \emph{{Large m(W), m(Z) Behavior of the O(alpha) Corrections to
  Semileptonic Processes Mediated by W}},
  \href{https://doi.org/10.1016/0550-3213(82)90303-0}{\emph{Nucl. Phys. B}
  {\bfseries 196} (1982) 83--92}.

\bibitem{vanRitbergen:1999fi}
T.~van Ritbergen and R.~G. Stuart, \emph{{On the precise determination of the
  Fermi coupling constant from the muon lifetime}},
  \href{https://doi.org/10.1016/S0550-3213(99)00572-6}{\emph{Nucl. Phys. B}
  {\bfseries 564} (2000) 343--390},
  [\href{https://arxiv.org/abs/hep-ph/9904240}{{\ttfamily hep-ph/9904240}}].

\bibitem{Chetyrkin:2000yt}
K.~G. Chetyrkin, J.~H. Kuhn and M.~Steinhauser, \emph{{RunDec: A Mathematica
  package for running and decoupling of the strong coupling and quark masses}},
  \href{https://doi.org/10.1016/S0010-4655(00)00155-7}{\emph{Comput. Phys.
  Commun.} {\bfseries 133} (2000) 43--65},
  [\href{https://arxiv.org/abs/hep-ph/0004189}{{\ttfamily hep-ph/0004189}}].

\bibitem{Jegerlehner:2001fb}
F.~Jegerlehner, M.~Y. Kalmykov and O.~Veretin, \emph{{MS versus pole masses of
  gauge bosons: Electroweak bosonic two loop corrections}},
  \href{https://doi.org/10.1016/S0550-3213(02)00613-2}{\emph{Nucl. Phys. B}
  {\bfseries 641} (2002) 285--326},
  [\href{https://arxiv.org/abs/hep-ph/0105304}{{\ttfamily hep-ph/0105304}}].

\bibitem{Jegerlehner:2002em}
F.~Jegerlehner, M.~Y. Kalmykov and O.~Veretin, \emph{{MS-bar versus pole masses
  of gauge bosons. 2. Two loop electroweak fermion corrections}},
  \href{https://doi.org/10.1016/S0550-3213(03)00177-9}{\emph{Nucl. Phys. B}
  {\bfseries 658} (2003) 49--112},
  [\href{https://arxiv.org/abs/hep-ph/0212319}{{\ttfamily hep-ph/0212319}}].

\bibitem{Arason:1991ic}
H.~Arason, D.~J. Castano, B.~Keszthelyi, S.~Mikaelian, E.~J. Piard, P.~Ramond
  et~al., \emph{{Renormalization group study of the standard model and its
  extensions. 1. The Standard model}},
  \href{https://doi.org/10.1103/PhysRevD.46.3945}{\emph{Phys. Rev. D}
  {\bfseries 46} (1992) 3945--3965}.

\bibitem{Brod:2020lhd}
J.~Brod and Z.~Polonsky, \emph{{Two-loop Beta Function for Complex Scalar
  Electroweak Multiplets}},
  \href{https://doi.org/10.1007/JHEP09(2020)158}{\emph{JHEP} {\bfseries 09}
  (2020) 158}, [\href{https://arxiv.org/abs/2007.13755}{{\ttfamily
  2007.13755}}].

\bibitem{Awramik:2003rn}
M.~Awramik, M.~Czakon, A.~Freitas and G.~Weiglein, \emph{{Precise prediction
  for the W-boson mass in the standard model}},
  \href{https://doi.org/10.1103/PhysRevD.69.053006}{\emph{Phys. Rev.}
  {\bfseries D69} (2004) 053006},
  [\href{https://arxiv.org/abs/hep-ph/0311148}{{\ttfamily hep-ph/0311148}}].

\bibitem{Buras:1993dy}
A.~J. Buras, M.~Jamin and M.~E. Lautenbacher, \emph{{The Anatomy of $\epsilon'
  / \epsilon$ beyond leading logarithms with improved hadronic matrix
  elements}}, \href{https://doi.org/10.1016/0550-3213(93)90535-W}{\emph{Nucl.
  Phys. B} {\bfseries 408} (1993) 209--285},
  [\href{https://arxiv.org/abs/hep-ph/9303284}{{\ttfamily hep-ph/9303284}}].

\bibitem{Chetyrkin:1997fm}
K.~G. Chetyrkin, M.~Misiak and M.~Munz, \emph{{Beta functions and anomalous
  dimensions up to three loops}},
  \href{https://doi.org/10.1016/S0550-3213(98)00122-9}{\emph{Nucl. Phys.}
  {\bfseries B518} (1998) 473--494},
  [\href{https://arxiv.org/abs/hep-ph/9711266}{{\ttfamily hep-ph/9711266}}].

\bibitem{BKPY}
J.~Brod, S.~Kvedaraite, Z.~Polonsky and A.~Youssef, \emph{{work in progress}},
  .

\bibitem{Binosi:2003yf}
D.~Binosi and L.~Theussl, \emph{{JaxoDraw: A Graphical user interface for
  drawing Feynman diagrams}},
  \href{https://doi.org/10.1016/j.cpc.2004.05.001}{\emph{Comput. Phys. Commun.}
  {\bfseries 161} (2004) 76--86},
  [\href{https://arxiv.org/abs/hep-ph/0309015}{{\ttfamily hep-ph/0309015}}].

\bibitem{Vermaseren:1994je}
J.~A.~M. Vermaseren, \emph{{Axodraw}},
  \href{https://doi.org/10.1016/0010-4655(94)90034-5}{\emph{Comput. Phys.
  Commun.} {\bfseries 83} (1994) 45--58}.

\bibitem{Gorbahn:2004my}
M.~Gorbahn and U.~Haisch, \emph{{Effective Hamiltonian for non-leptonic
  $|\Delta F| = 1$ decays at NNLO in QCD}},
  \href{https://doi.org/10.1016/j.nuclphysb.2005.01.047}{\emph{Nucl. Phys. B}
  {\bfseries 713} (2005) 291--332},
  [\href{https://arxiv.org/abs/hep-ph/0411071}{{\ttfamily hep-ph/0411071}}].

\bibitem{Buras:1991jm}
A.~J. Buras, M.~Jamin, M.~E. Lautenbacher and P.~H. Weisz, \emph{{Effective
  Hamiltonians for $\Delta S = 1$ and $\Delta B = 1$ nonleptonic decays beyond
  the leading logarithmic approximation}},
  \href{https://doi.org/10.1016/0550-3213(92)90345-C}{\emph{Nucl. Phys. B}
  {\bfseries 370} (1992) 69--104}.

\bibitem{Herrlich:1994kh}
S.~Herrlich and U.~Nierste, \emph{{Evanescent operators, scheme dependences and
  double insertions}},
  \href{https://doi.org/10.1016/0550-3213(95)00474-7}{\emph{Nucl. Phys. B}
  {\bfseries 455} (1995) 39--58},
  [\href{https://arxiv.org/abs/hep-ph/9412375}{{\ttfamily hep-ph/9412375}}].

\end{thebibliography}\endgroup

\end{document}